\documentclass[5p,authoryear]{elsarticle}

\usepackage{amssymb}
\usepackage{graphicx}

\def \curlB {\vec{\nabla}\times \vec{B}}

\journal{Computer Physics Communications}

\begin{document}

\begin{frontmatter}

\title{A new code for the Hall-driven magnetic evolution of neutron stars}

\author{D.~Vigan\`o}
 \ead{daniele.vigano@ua.es} 
\author{J.A.~Pons}
\author{J.A.~Miralles}

\address{Departament de F\'{\i}sica Aplicada, Universitat d'Alacant, Ap. Correus 99, 03080 Alacant, Spain}

\begin{abstract}
Over the past decade, the numerical modeling of the magnetic field evolution in astrophysical scenarios has become an increasingly important field. In the crystallized crust of neutron stars the evolution of the magnetic field is governed by the Hall induction equation. In this equation the relative contribution of the two terms (Hall term and Ohmic dissipation) varies depending on the local conditions of temperature and magnetic field strength. This results in the transition from the purely parabolic character of the equations to the hyperbolic regime as the magnetic Reynolds number increases, which presents severe numerical problems. Up to now, most attempts to study this problem were based on spectral methods, but they failed in representing the transition to large magnetic Reynolds numbers. We present a new code based on upwind finite differences techniques that can handle situations with arbitrary low magnetic diffusivity and it is suitable for studying the formation of sharp current sheets during the evolution. The code is thoroughly tested in different limits and used to illustrate the evolution of the crustal magnetic field in a neutron star in some representative cases. Our code, coupled to cooling codes, can be used to perform long-term simulations of the magneto-thermal evolution of neutron stars.
\end{abstract}


\end{frontmatter}

\section{Introduction}

The magnetic field (MF) is a key agent to explain many of the observed properties of neutron stars (NSs). In order to explain NS rotational energy losses by dipole radiation, a simple back-of-the-envelope estimate gives that NSs must have an external dipolar component ranging from $10^{10}$ to $10^{15}$ G,  but little is known about its internal geometry and evolution. In the standard NS evolution model, a few hours, or at most a few days after birth, a solid crust about 1 km thick (compared to the typical 10 km radius) is formed. Under these conditions, ions in the lattice have very restricted mobility and conduction is governed by the electrons. At the large densities of a neutron star crust ($10^{10}-10^{14}$ g cm$^{-3}$), matter is completely pressure-ionized and electrons can flow trough the lattice. The equation describing the evolution of such system is the generalized Hall induction equation for a one-component plasma (electrons). This limit is also known as Electron MHD (EMHD), and describes a one-component charged fluid which density does not vary with time.
 
The evolution of the crustal MF is thus regulated by Ohmic dissipation, with timescales of $10^6$ yr, and the Hall term. The latter causes a transfer of energy to smaller scales and the coupling between toroidal and poloidal field components, on timescales of $10^3-10^4$ yr. The Hall-driven evolution is probably at the origin of the observed strong activity in young {\it magnetars}, i.e., NSs with ultra-strong magnetic fields. Therefore, the interest in modeling the internal evolution of the MF is continuously growing \citep{pons07,hoyos08,pons09,shabaltas12}. The MF in the crust of NSs also directly affects the microphysical processes that govern the thermal evolution of the crust. For this reason, previous works \citep{aguilera08,pons07,pons09} were directed toward a fully coupled magneto-thermal evolution of the NS crust. \cite{aguilera08} developed a 2D (axial symmetry) cooling code taking into account the interplay between temperature and MF. However the decay of the latter was simulated by a phenomenological, homologous analytical formula, without solving the induction equation because of numerical limitations in the treatment of the Hall term. One of the important parameters is the resistivity, which strongly depends on the local temperature. For typical temperatures (a few $10^8$ K) in a middle-age NS of $10^3-10^5$ yr, the Hall term dominates over Ohmic dissipation when the MF is $> 10^{14}$ G. However, as the star cools down and the Ohmic timescale becomes very long (billions of years), the Hall term may dominate even for much lower values of the MF strength, which in turn requires a better numerical approach.

The numerical challenge is to be able to follow the evolution of such system, in which important parameters (density, resistivity) vary several orders of magnitude across the spatial domain, but also during the temporal evolution. A multi-purpose code must be able to work in both the purely diffusive regime and in the limit of large magnetic Reynolds number when the Hall term dominates. It must also be stable to follow the evolution for many diffusion timescales (up to hundreds of Hall timescales).  Historically, spectral methods had been used to solve the Hall induction equation in simplified constant density layers \citep{hollerbach02}, but such attempts were always restricted to a magnetic Reynolds number not exceeding 200, since fully spectral codes systematically have unsurmountable problems dealing with structures where discontinuities or very large gradients of the variables appear, which is a natural consequence of the equations. \cite{pons07} presented a code solving the Hall induction equation in a realistic crust using an alternative approach (spectral in angles but finite differences in the radial direction). This was a significant
improvement to previous work, but still could not work in the limit of vanishing electrical resistivity. For this reason, the only long-term fully coupled 2D magneto-thermal evolution simulations available up to now \citep{pons09} were restricted to the purely diffusive case. 

In this paper we present a new code based on upwind, finite difference schemes that can handle the Hall term in the induction equation for vanishing physical resistivity. Preliminary results using the code described in this paper, and confirming the occurrence of the Hall instability in neutron stars were presented in \cite{pons10}.
The shock-capturing character of the method employed ensures that the formation of current sheets is properly modeled. The code can follow the evolution of complex geometries with large gradients and discontinuities overcoming intrinsic problems of previous studies. The paper is organized as follows. In \S \ref{sec_induction} we describe the generalized induction equation, and discuss the importance of the resistive and Hall terms in the system. In \S \ref{sec_code} we present the numerical code, describing the grid and the scheme used to solve the induction equation. In \S \ref{sec_test} we present the battery of tests we have performed, and show some representative models. In \S \ref{tor_sec} we consider the evolution of a purely toroidal field in a realistic NS crust, with particular attention to the formation of 
discontinuities. In \S \ref{all_sec} we present a general case with a realistic mixture of poloidal and toroidal MF. 
In \S \ref{sec_conclusions} we summarize our findings and discuss future applications of physical interest.

\section{The EMHD limit and the Hall induction equation}\label{sec_induction}

In the EMHD approximation, there is the implicit assumption that the timescale of 
variation of electromagnetic field is much larger than the timescale of collisions inside plasma. Therefore 
we can neglect displacement currents in Ampere's law and the current density is simply
\begin{equation}\label{current_mhd}
 \vec{J} =\frac{c}{4\pi}\curlB~.
\end{equation}
In a one-component plasma (in this case, electrons), the velocity is not an independent variable, but instead it is related to the current and charge densities by 
\begin{equation}\label{hall_velocity}
 \vec{v}=-\frac{\vec{J}}{en_e} = - \frac{c}{4\pi e n_e} \curlB ~,
\end{equation}
where $n_e$ is the electron number density, $e$ the elementary charge, and $c$ the speed of light.

For a conducting fluid, moving with velocity $\vec{v}$, and in the presence of a MF, the electric field ($\vec{E}$) is related to the current density through the generalized Ohm's law 
\begin{equation}\label{ohm_general}
\vec{J} = \sigma\left(\vec{E}+\frac{\vec{v}}{c}\times\vec{B}\right)~,
\end{equation}
where $\sigma$ is the electrical conductivity.

The evolution of this system is only governed by the Hall induction equation (e.g. \cite{pons07} and references therein), 
\begin{equation}\label{induction_Hall}
\frac{\partial \vec{B}}{\partial t} = - \vec{\nabla}\times \left\{ 
\eta \curlB + \frac{c}{4\pi e n_e} (\curlB) \times\vec{B} \right\}
\end{equation}
which can be derived from the induction equation 
\begin{equation}\label{induction_general}
 \frac{\partial \vec{B}}{\partial t} =  -c \vec{\nabla}\times \vec{E}~,
\end{equation}
where, from Eqs.~(\ref{hall_velocity}) and (\ref{ohm_general}), the electric field is given by
\begin{equation}\label{efield_hall}
c \vec{E}= \eta \curlB + \frac{c}{4\pi e n_e} (\curlB) \times\vec{B}.
\end{equation}
Since the magnetic diffusivity, $\eta=c^2/4\pi\sigma$, depends on the temperature, the magnetic and thermal evolution 
in the crust of a NS are coupled.
The first term on the right-hand side of Eq.~(\ref{induction_Hall}) is responsible for the Ohmic decay. 
The second term is the Hall term, and its non-linear character is in the root of all difficulties
one finds when numerical methods are used to solve the Hall induction equation.

To compare the relative importance of the two terms (Ohmic dissipation and Hall), 
we can write the Hall induction equation (\ref{induction_Hall}) as follows:
\begin{equation}
\frac{\partial \vec{B}}{\partial t} = -\vec{\nabla}\times \left( \eta \left\{\curlB + 
{\cal R}_m {(\curlB)\times\vec{b}} \right\}\right)\\
\end{equation}
where $\vec{b}$ is the unit vector in the direction of the MF and ${\cal R}_m=cB/4\pi en_e \eta$ is the magnetic Reynolds number. ${\cal R}_m$ is an indicator of the relative importance between the two terms on the right-hand side. 

We define the characteristic Ohmic dissipation timescale 
\begin{equation}
\tau_{d}=L^2/\eta~,
\end{equation} 
and Hall timescale
\begin{equation}
\tau_{h}= 4\pi e n_e L^2/c B = \tau_{d}/{\cal R}_m~,
\end{equation} 
where $L$ is a typical length scale. With these definitions, ${\cal R}_m$ is the ratio between the dissipation and Hall timescales. It also corresponds to the so-called ``magnetization parameter'' ($\omega_B\tau_{e}$), usually found in the literature of plasma physics, where $\omega_B$ is the cyclotron frequency and $\tau_{e}$ the electron relaxation time.

\section{The numerical code}\label{sec_code}

In order to understand the dynamical evolution of the system and to design a successful numerical algorithm, it is important
to know the mathematical character of the equations and to identify the wave modes.
The magnetic Reynolds number determines the relative importance of the Hall term and Ohmic dissipation, and defines
the transition from a purely parabolic equation (${\cal R}_m \ll 1$), to a hyperbolic regime (${\cal R}_m \gg 1$).
In the case of Hall-dominated evolution, the problem becomes a ``doubly'' constrained multidimensional 
advection equation because, in addition to the $\vec{\nabla}\cdot\vec{B}=0$ constraint, 
the velocity field is proportional to the current (Eq.~\ref{hall_velocity}), i.e., to the derivatives of the field. 
This implies non-linearity and coupling between poloidal and toroidal components. An initially pure toroidal field will not develop a poloidal component, but any poloidal field will induce the formation of toroidal components. 
When both components are present, magnetic energy can be transferred between them.
Although the Hall term itself conserves energy, the plausible creation of small-scale structures
may accelerate the Ohmic dissipation. 

The Hall term also introduces two wave modes into the system. \cite{huba03} has shown that,
in a constant density medium, the only modes of the Hall EMHD equation are the {\it whistler or helicon waves}, which are
transverse field perturbations propagating along the field lines. In presence of a charge density gradient, 
additional {\it Hall drift waves} appear. These are transverse modes that
propagate in the $\vec{B} \times \vec{\nabla} n_e$ direction. 
It has also been noted that the presence of charge density gradients results in a Burgers-like
term \citep{vainshtein00}. Furthermore, even in the constant density case but without planar symmetry, 
the evolution of the toroidal component of the MF  
also contains a quadratic term that resembles the Burgers equation \citep{pons07} with
a coefficient dependent on the distance to the axis.
A natural outcome of the presence of Burgers-like terms is the formation of ``shocks'' (current sheets). 
The evolution for the poloidal field is more complicated, as it includes higher order derivatives in the non-linear terms. 
All the above issues must be taken into account when an efficient and robust numerical method is
designed to solve the Hall induction equation in the limit of strong magnetization. To overcome
all previous limitations found when techniques based on spectral methods have been used,
we decided to use well-known upwind, conservative schemes. These methods are of very general use,
from problems involving the simple 1D Burgers equation to high resolution shock capturing schemes 
very successfully used in MHD problems \citep{anton06,giacomazzo07,cerdaduran08}.

\subsection{The numerical grid: staggered grids}
Staggered grids \citep{yee66} are commonly used with finite difference time domain methods \citep{taflove75} to solve the Maxwell's equations. The MF components  are defined at the center, and normal to each face of the cell, while the electric field is defined along the edges of the cell. With these definitions, the  $\vec{\nabla}\cdot\vec{B}=0$ condition can be satisfied to machine error (see below). The components of $\curlB$, as the electric field, are also defined along the edges of the cell. Stokes' theorem allows us to write the curl components in terms of the MF components defined at the cell faces.

\begin{figure}
 \centering
\includegraphics[width=.25\textwidth]{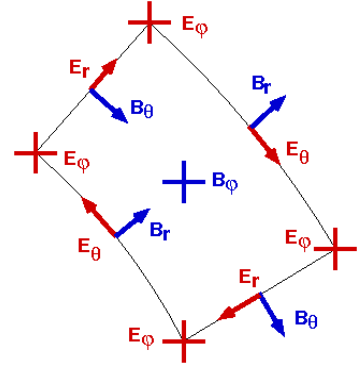}
\caption{Location of the variables on a staggered grid in spherical coordinates and for the axisymmetric case.
Solid lines delimit the edges of the surface $\Sigma_\varphi$.} 
 \label{fig_staggered}
\end{figure}

As an example, we show in Fig.~\ref{fig_staggered} the location of the variables in a numerical cell in spherical coordinates $(r, \theta, \varphi)$ and assuming axial symmetry. In this case our grid maps a meridional section of the star. In this paper we will restrict to 2D simulations, but we will present tests and results in both Cartesian and spherical coordinates.

\subsection{Conservation form and divergence-preserving schemes.}

One important feature of numerical schemes designed for hyperbolic problems is the formulation of the equations in conservative form. This ensures a correct propagation of waves and the preservation (by construction) of the divergence constraint.  Integrating over the surface of a cell face, normal to the $\alpha$ direction, and applying Stokes theorem to Eq.~(\ref{induction_general}), the space-discretized equation for the magnetic flux $\Phi_\alpha$ can be written as
\begin{eqnarray}\label{phialpha}
\frac{1}{c}\frac{d \Phi_\alpha}{d t}  =  - \sum_k E_k l_k ~.
\end{eqnarray}
The circulation of the $\vec{E}$-field around a face boundary has been approximated by the sum $\sum_k E_k l_k$, where $E_k$ are the central values of the electric field on the edges, $l_k$ is the length of the edge, and $k$ runs over the four edges of the face. The normal components of the MF to a given face are assumed to be average values over the surface, $B_\alpha=\Phi_\alpha/\Sigma_\alpha$, where $\Sigma_\alpha$ is the face area.

Making use of Gauss' theorem, the numerical divergence can be evaluated, for each cell with volume $\Delta V$, as follows:
\begin{equation}\label{divb_staggered}
  \vec{\nabla}\cdot\vec{B}=\frac{1}{\Delta V}\sum_\alpha{(\Phi_\alpha^{+}-\Phi_\alpha^{-})}~,\nonumber
\end{equation}
where the plus and minus superscripts indicate the fluxes through the two opposite faces in a given direction $\alpha$ (in the 2D example of Fig.~\ref{fig_staggered}, $\alpha=r,\theta$). With this definition, the divergence-preserving character of the methods using the conservation form and advancing in time magnetic fluxes, instead of MF components, becomes evident: taking the time derivative of Eq.~(\ref{divb_staggered}), and using Eq.~(\ref{phialpha}), every edge contributes twice with a different sign and cancels out.

\subsection{Cell reconstruction}

The original Godunov's method is well known for its ability to capture discontinuous solutions, but it is only first-order accurate. This method can be easily extended to give second-order spatial accuracy on smooth solutions, but still avoiding non-physical oscillations near discontinuities. To achieve higher order accuracy, we can use a reconstruction procedure that improves the piecewise constant approximation. The simplest choice is a piecewise linear function in each cell. A very popular choice for the slopes of the linear reconstructed function is the {\it monotonized central-difference limiter} (MC), proposed by \cite{vanleer77}. Given three consecutive points $x_{i-1},x_i,x_{i+1}$ on a numerical grid, and the numerical values of the function $f_{i-1},f_i,f_{i+1}$, the reconstructed function within the cell $i$ is given by $f(x)=f(x_i)+ \alpha (x-x_i)$, where the slope is
$$\alpha = {\rm minmod}\left( \frac{f_{i+1}-f_{i-1}}{x_{i+1}-x_{i-1}},2\frac{f_{i+1}-f_{i}}{x_{i+1}-x_{i}},
2\frac{f_{i}-f_{i-1}}{x_{i}-x_{i-1}}\right).$$
The ${\rm minmod}$ function of three arguments is defined by
\begin{equation}
{\rm minmod}(a,b,c) = \left\{ 
\begin{array}{cc}
{\rm min}(a,b,c) & {\rm if} ~a,b,c>0 ;\\
{\rm max}(a,b,c) & {\rm if} ~a,b,c<0 ;\\
0 & {\rm otherwise} .
\end{array}
\right.\nonumber
\end{equation}
We reconstruct the MF circulation elements, $C_k= B_k l_k$. From this,
we directly obtain the components of $(\curlB)$ by means of Stokes' theorem and, when needed, 
we recover $B_x$ dividing the reconstructed circulation by the local length element.

\begin{figure}
 \centering
 \includegraphics[width=.45\textwidth]{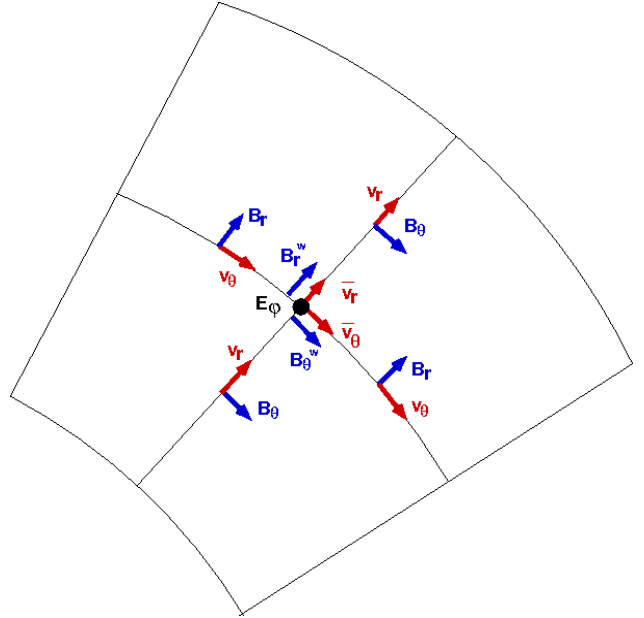}
\caption{Illustration of the procedure of calculation of the electric field: location of the 
components of velocity (red arrows) and MF (blue) involved in the definition of the Hall term of $E_\varphi$ (black dot).}
 \label{fig_ewind}
\end{figure}

From Eqs.~(\ref{hall_velocity}) and (\ref{efield_hall}), the electric field is:
\begin{eqnarray}\label{efield_v}
\vec{E} = \frac{4 \pi}{c^2}\eta  \vec{J} -\frac{1}{c}\vec{v} \times \vec{B}
\end{eqnarray}
Current and electric field components are always defined at the same location, but the Hall term includes products of tangential components of the MF and velocity not always defined at the same edges. For such terms, we evaluate the needed components of the velocity by linear interpolation of the closest neighbor values and we take the {\it upwind} 
components $\vec{B}^w$ of the reconstructed value of MF at each interface. For example, in the axisymmetric case and considering the evolution of the poloidal components, the contributions of $E_r$ and $E_\theta$ to the circulation cancel out and we only need to evaluate the contribution of $E_\varphi$, which is calculated by
\begin{eqnarray}\label{efield_wind}
{E_\varphi} =\frac{4 \pi}{c^2}\eta  {J_\varphi}  - \frac{1}{c} \left( \bar{v}_r {B_\theta}^w -  \bar{v}_\theta {B_r}^w \right) ~.
\end{eqnarray}
In Fig.~\ref{fig_ewind} we explicitly show the location of $E_\varphi$ (black point) and the location on the staggered grid of the quantities needed for its evaluation. The average values of $\bar{v}_r$ and $\bar{v}_\theta$ are calculated taking 
the average of the two closest neighbors. In the example, both $\bar{v}_r$ and $\bar{v}_\theta$ are positive, therefore the 
{\it upwind} values of  $B_r^w$ and $B_\theta^w$ are the reconstructed values from the bottom and left sides,
respectively.

\subsection{Time advance and Courant condition}

We use an explicit, first-order time advance method with some intermediate corrections to improve the stability of the scheme. In explicit algorithms, the time step is limited by the Courant condition, that avoids that any wave travels more than one cell length on each time step. Since we want to evolve the system on long (Ohmic) timescales, the Courant condition is particularly restrictive when ${\cal R}_m\gg 1$. At each time step, we estimate the Courant time ($t_c$) by 
\begin{equation}\label{estimate_courant_hall}
 t_c = \mbox{min} \left(\frac{4\pi e n_e ~ \Delta l }{c|\vec{\nabla} \times \vec{B}|}\right)_{i,j}, \end{equation}
where $\Delta l$ is the minimum length of the cell edges in any direction and $(i,j)$ denote the numerical cells. We use a time step
\begin{equation}\label{timestep}
 \Delta t=k_c t_c,
\end{equation}
where $k_c$ is a factor $\le 1$. The time step can vary by orders of magnitude during the evolution, becoming very small when the Hall term dominates and, in particular, where locally strong current sheets are formed. In a typical situation, the Ohmic timescale is of the order of $10^6$ yr, the Hall timescale is $10^3-10^4$ yr, and the Courant condition limits the time step between $10^{-3}$ and 1 yr for a typical number of grid points of $50\times 50$.

After calculating all the line integrals of $\vec{E}$ along all edges, the time advance proceeds in two different steps. Hereafter we denote by the subscript $t$ the component of the MF in the symmetry direction (the toroidal component in spherical coordinates, with axial symmetry), and by the subscript $p$ the other two components (poloidal). First we advance the $t$ component of the MF, with a particular treatment of the quadratic term in $B_t$ (see \S~\ref{sec_burgers}). With the updated value of this component, the electric field components are recalculated and then used to advance the $p$ components of the MF. 
Schematically, the sequence of the time advance from $t_n$ to $t_{n+1}=t_n+\Delta t$ is the following:
\begin{itemize}
\item
starting from $\vec{B}^n$, all currents and electric field components are calculated \\ 
$\vec{B}^n\rightarrow \vec{J}^n\rightarrow \vec{E}^n$;
\item
$\vec{B}_t^{n}$ is updated: $\vec{E}^n \rightarrow \vec{B}_t^{n+1}$;
\item
the new values $\vec{B}_t^{n+1}$ are used to calculate the modified current components and $\vec{E}_t$: \\
$\vec{B}_t^{n+1} \rightarrow \vec{J}_p^\star \rightarrow \vec{E}_t^\star$;
\item
finally, we use the values of $\vec{E}_t^\star$ to update the remaining MF components \\
$\vec{E}_t^\star \rightarrow \vec{B}_p^{n+1}$.
\end{itemize}

This two-step advance favors the stability of the method, as already pointed out for a 3D problem in Cartesian coordinates by \cite{osullivan06}. \cite{toth08} further discussed that the two-stage formulation is equivalent to introduce a fourth-order hyper-resistivity term. In our case, since the $t$-component is advanced explicitly, the hyper-resistive correction
only acts on the evolution of the $p$-components. The correction introduced by the intermediate step is:
\begin{equation}
\delta\vec{B}_p^{n+1}=-c\Delta t\,\vec{\nabla}\times\delta\vec{E}_t
\end{equation}
where $\delta\vec{E}_t = \vec{E}_t^* - \vec{E}_t^n$. After some algebra, one can show that
$$\delta\vec{B}_p^{n+1}=(c \Delta t)^2\,\vec{\nabla}\times\left\{ \frac{c}{4\pi e n_e}  \{[\vec{\nabla}\times(\vec{\nabla}\times\{\eta(\curlB_t^n)  \right. $$
$$\left. +\frac{c}{4 \pi e n_e}  [(\curlB_p^n)\times\vec{B}_p^n - (\curlB_t^n)\times\vec{B}_t^n]\})]\times\vec{B}_p^n\} \right\}$$
From this expression, we can see that the additional correction given by $\delta\vec{E}_t$ contains third and fourth-order spatial derivatives and scales with $(\Delta t)^2$, which is characteristic of hyper-resistive terms. We have found a significant improvement in the stability of the method when comparing a fully explicit algorithm with the two-steps method.
The method is always stable, if a sufficiently small value of $k_c$ is used (typically $10^{-2}$-$10^{-1}$).

\subsection{Energy balance}

The magnetic energy balance equation for Hall EMHD can be expressed as:
\begin{equation}\label{en_cons}
 \frac{\partial}{\partial t}\left(\frac{B^2}{8\pi}\right)= - Q_j - \vec{\nabla}\cdot \vec{S}
\end{equation}
where $Q_j=4\pi\eta J^2/c^2$ is the Joule dissipation rate and $\vec{S}=c\vec{E}\times\vec{B}/4\pi$ is the Poynting vector. 
During the evolution, the magnetic energy in a cell can only vary due to local Ohmic dissipation and by interchange
between neighbor cells (Poynting flux).

Integrating Eq.~(\ref{en_cons}) over the whole volume of the numerical domain, we obtain the balance between the time variation of the total magnetic energy ${\cal E}_b=\int_V (B^2/8\pi) dV$, the total Joule dissipation rate ${\cal Q}_{tot}=\int_V Q_j dV$, and the Poynting flux through the boundaries ${\cal S}_{tot}=\oint_{\partial V} \vec{S}\cdot d\vec{\Sigma}$:
\begin{equation}\label{integrated_balance}
 \frac{d}{d t}{\cal E}_b + {\cal Q}_{tot} + {\cal S}_{tot}=0~.
\end{equation}
A necessary test for any numerical code is to check the instantaneous (local and global) energy balance given by 
Eqs.~(\ref{en_cons}) and (\ref{integrated_balance}). 
Any type of numerical instability usually results in the violation of the energy 
conservation. Therefore a careful monitoring of the energy balance is a powerful diagnostic test.

\section{Numerical tests.}\label{sec_test}
We have performed a battery of tests, with special attention to the conservation of the total energy 
and the numerical viscosity of the method employed.
In order to understand the different aspects of Hall MHD, in this section we show some illustrative examples of 2D Cartesian tests, reproducing whistler and Hall drift waves. Finally, we will discuss the decay of Ohmic modes in a constant density sphere to compare against the analytic solutions for the purely resistive (${\cal R}_m=0$) case. 

\subsection{2D tests in Cartesian coordinates.}

\begin{figure}[ht!]
 \centering
 \includegraphics[width=.4\textwidth]{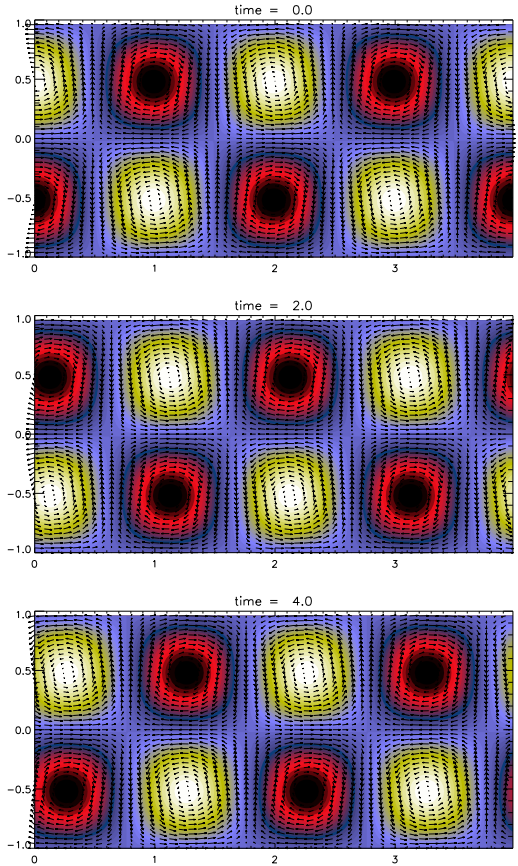}
\caption{Evolution of the initial configuration defined by Eqs.~(\ref{inimod1}) with $B_0=10^3~B_1$ and $k_x L=\pi$ at three different times (in units of $\tau_0$). Arrows show the perturbed $B_x$ (subtracting $B_0$), and $B_z$-components, while the color scale represents the $B_y$ component (red/black positive, yellow/white negative).} 

\label{fig_whistler}
\end{figure}

\subsubsection{Whistler waves.}
The first test we have performed is to follow the correct propagation of  whistler (or helicon) waves in a two-dimensional slab, extending from $z=-L$ to $z=+L$ in the vertical direction. We impose periodic boundary conditions in the $x$-direction. All variables are independent of the $y$-coordinate. In the case of constant density, $n_e=n_0$, and $\eta=0$ (i.e. infinite magnetic Reynolds number), the only modes present in the system are whistler waves. We consider the Hall induction equation for the following initial MF
\begin{eqnarray}
\label{inimod1}
B_x&=& B_0 + B_1\cos(k_x z)\cos(k_x x); \nonumber \\
B_y&=& \sqrt{2}B_1 \sin(k_x z)\cos(k_x x) ; \nonumber \\
B_z&=& B_1\sin(k_x z)\sin(k_x x);
\end{eqnarray}
where $k_x = n \pi /L$, $n=1,2,...$, and $B_1\ll B_0$. We define the reference Hall timescale as
\begin{equation}\label{tau0_wave}
\tau_0=\frac{4\pi e n_0 L^2}{c B_0}~.
\end{equation}
This case admits a pure wave solution confined in the vertical direction and traveling in the $x$-direction with speed 
\begin{equation}\label{vel_whistler}
v_w=-\frac{c}{4\pi e n_0}\sqrt{2}~k_x B_0=-\sqrt{2}\frac{L^2 k_x}{\tau_0}~,
\end{equation}

The evolution of the initial configuration defined by Eqs.~(\ref{inimod1}) with $B_0=10^3~B_1$, and $k_x L=\pi$ at three different times (in units of $\tau_0$) is shown in Fig.~\ref{fig_whistler}, and for a $200 \times 50$ grid. The travel time to cross over the whole domain is $t=0.9~ \tau_0$, thus the perturbations have crossed through the horizontal domain several times, without apparent dissipation or shape change. The code runs for hundreds of Hall timescales without any indication of instabilities, despite the fact that electrical resistivity is set to zero. The measured speed of the whistler waves is 0.9992 the analytical value for this particular resolution. We have also checked that the correct scaling of the propagation velocity (linear with $k_x$ and $B_0$) is recovered.

\begin{figure}[!t]
 \centering
 \includegraphics[width=.4\textwidth]{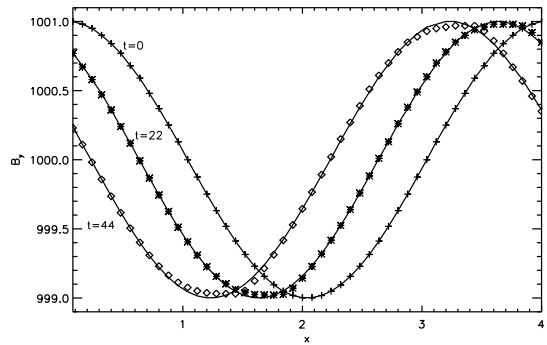}
\caption{Horizontal section ($z=0$) of the evolution of the initial configuration defined by Eq.~(\ref{inimod2}) with $B_0=10^3~B_1$, $L=1$ and $k_x L=\pi/2$  at $t=0,22,$ and 44 (in units of $\tau_0$). 
The perturbation crosses over the entire domain in $20 ~\tau_0$.} 
\label{fig_drift}
\end{figure}

\subsubsection{Hall drift waves.}
As a second test in Cartesian coordinates, we set up the following initial configuration:
\begin{eqnarray}\label{inimod2}
B_x&=& 0; \nonumber \\
B_y&=& B_0 + B_1\cos(k_x x) ; \nonumber \\
B_z&=& 0;
\end{eqnarray}
on a stratified background in the $z$-direction with 
\begin{equation}\label{n_wave}
n_e(z)= \frac{n_0}{1+\beta z}~,
\end{equation}
where $n_0$ is a reference density to which we associate the same Hall timescale $\tau_0$ defined in Eq.~(\ref{tau0_wave}), $z\in[-L,L]$, and $\beta$ is a parameter with dimensions of inverse length. We impose again periodic boundary conditions in the $x$-direction. In the $z$-direction, we copy the values of the MF in the first and last cells ($z=\mp L$) on the first neighbor ghost cell, to simulate an infinite domain. For small perturbations ($B_1\ll B_0$), this configuration must induce Hall drift waves traveling in the $x$-direction with speed

\begin{equation}\label{vel_hd}
  v_{hd}=-\frac{\beta L^2}{\tau_0}
\end{equation}

For the particular model shown in Fig.~\ref{fig_drift}, with $B_0=10^3~B_1, k_x L=\pi/2$, and $\beta L=0.2$, this gives a horizontal drift velocity of $v_{hd}=-0.2 L/\tau_0$. We show the evolution of $B_y$ at three different times, when the perturbation has crossed over the numerical domain 1.1 and 2.2 times, respectively. The resolution used is $50\times50$. For the Hall drift modes, the propagation velocity scales linearly with $B_0$ and the 
gradient of $n_e^{-1}$, but it is independent of the wavenumber of the perturbation. All these properties are correctly reproduced.

\subsubsection{The nonlinear regime and Burgers flows.}\label{sec_burgers}

\begin{figure}
 \centering
 \includegraphics[width=.4\textwidth]{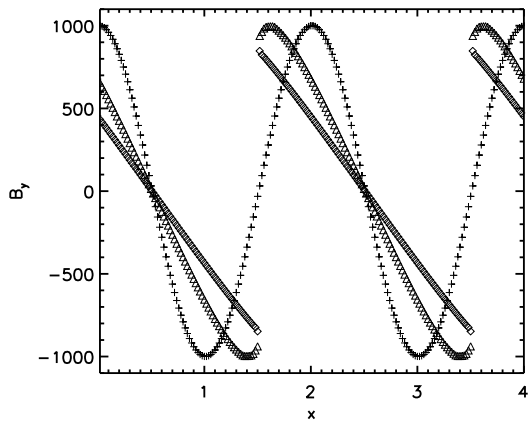}
\caption{Horizontal section of the evolution of the initial configuration defined by Eq.~(\ref{inimod3}) with $B_0=10^3$ and $k_x L=\pi$ at $t=0,2,$ and 4 (in units of $\tau_0$). The shock forms at $t=2$. The classical sawtooth shape developed during the evolution of the Burgers equation is evident.} 
\label{fig_burg1}
\end{figure}

With the two previous tests, we can conclude that the code is able to accurately propagate the two fundamental modes with the correct speeds. However, the Hall drift modes are a valid solution only in the linear regime. Let us consider more carefully the evolution of the $B_y$ component in a medium stratified in the $z$-direction. Assuming that $B_x=B_z=0$, it can be readily derived that the governing equation reduces to
\begin{equation}
 \frac{\partial B_y}{\partial t} + g(z) {B_y}\frac{\partial B_y}{\partial x}=0,
\end{equation}
where $g(z)=-\frac{d}{dz}(\frac{c}{4\pi e n_e})$. This is nothing but the familiar Burgers equation in the $x$-direction, with a coefficient that depends on the $z$ coordinate. Thus, the evolution of the $B_y$ component of the MF is governed by independent Burgers-like equations with different propagation speeds at different heights. The solution of such equation in one dimension is well known and has been studied for decades. As a last test in Cartesian coordinates we consider the following initial configuration:
\begin{eqnarray}
\label{inimod3}
B_x&=& 0; \nonumber \\
B_y&=& B_0 \cos(k_x x) ; \nonumber \\
B_z&=& 0;
\end{eqnarray}
on the same stratified background as in the previous subsection, Eq.~(\ref{n_wave}), which gives simply $g(z)=-\beta L^2/\tau_0 B_0$ and allows the direct comparison to the solution of the Burgers equation in one dimension.

\begin{figure*}[!ht]
 \centering
 \includegraphics[width=.95\textwidth]{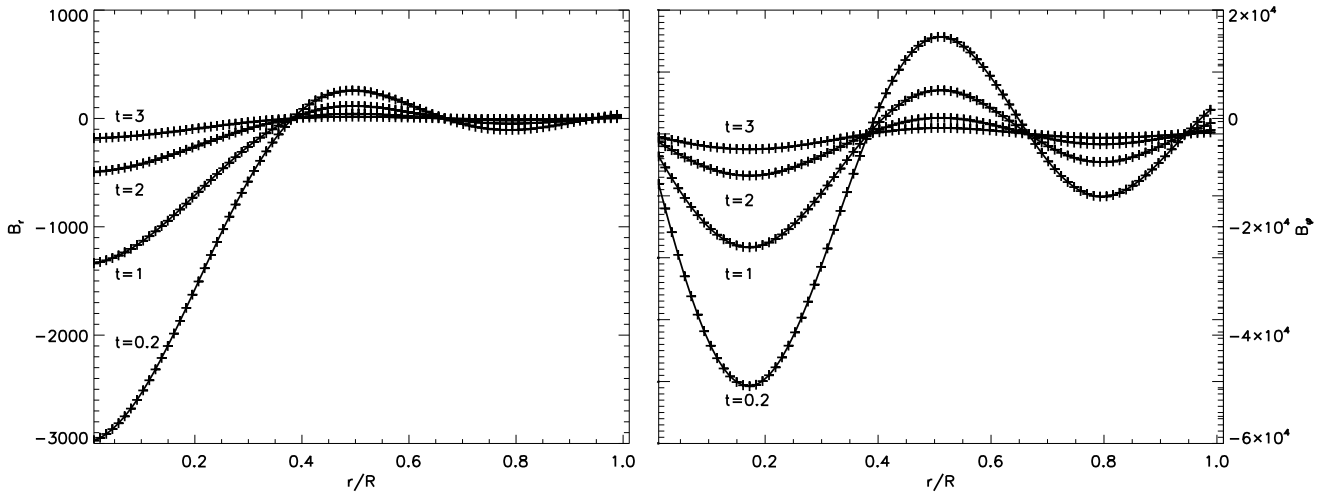}
\caption{Evolution of the purely Ohmic modes (Eq.~\ref{bessel}), at $t=0.2, 1, 2,$ and 3 diffusion times ($\tau_d$) for the model with $\mu R=11.5$ and a resolution of  $88\times60$ equally spaced grid points. In the figure we compare the 
analytical (lines) and numerical (crosses) radial profiles of $B_r$ (left) and $B_\varphi$ (right) at $\theta=\pi/3$.}
\label{fig_bessel_decay}
\end{figure*}

Making use of the well-known numerical techniques applied to the Burgers equation, in our code
we give an special treatment to the quadratic term in $B_y$ in the evolution equation for that component
(we will proceed analogously for the $\varphi$-component in spherical coordinates). 
The key issue is to write the Burgers-like term in conservation form:
\begin{equation}
 \frac{\partial B_y}{\partial t} + g(z)\frac{\partial (B_y^2/2)}{\partial x}=0,
\end{equation}
which can be discretized by an upwind conservative method
\begin{equation}
\frac{d B_y^j}{dt}= - \frac{\left( \hat{F}^{j+1/2}- \hat{F}^{j-1/2}\right)}{\Delta x},
\end{equation}
where the problem is reduced to calculate the upwind numerical fluxes at the interfaces $\hat{F}^{j\pm1/2}$.
In this case the wave velocity determining the upwind direction is given by $\lambda = g(z)B_y$ and the flux
$\hat{F}=g(z) B_y^2/2$.
Using methods in flux-conservative form is particularly important when solving problems with shocks or other discontinuities
 (e.g. \cite{Toro}),
as a non-conservative method may give numerical solutions that look reasonable but are entirely wrong (incorrect
propagation speed of discontinuous solutions). 
The particular treatment of the time advance of this term (different from the general method described in Section 3.2)
has no impact on the divergence-preserving character of the method, since this is the component in the 
symmetry direction of the problem (the same argument applies for the toroidal field in  axial symmetry). 

In Fig.~(\ref{fig_burg1}) we show snapshots of the evolution of the initial conditions (\ref{inimod3}) 
with $k_x L=\pi$, $B_0=10^3$, and $\beta L=0.2$. It follows the typical Burgers evolution. The wave breaking and
the formation of a shock at $t=2 \tau_0$ is clearly captured, as would happen in the solution of the inviscid Burgers equation.
We must stress again that this test is done with zero physical resistivity, i.e., in the limit ${\cal R}_m \rightarrow \infty$,
which is not reachable by spectral methods or standard centered difference schemes in non-conservative form.

We should also stress that the evolution of the energy spectrum of a Burgers-like problem leads to the scaling $k^{-2}$, as shown for example in \cite{tran10}. This is exactly the expected scaling, first discussed in the context of Hall MHD in neutron stars in \cite{goldreich92}, who pointed out the analogies and differences with turbulent velocity fields. The $k^{-2}$ scaling for the power spectrum has been discussed in the past in terms of a Hall cascade, transferring energy from larger scale to smaller scale modes, and there 
is some open debate about its interpretation as a global turbulent cascade or a local steepening of some magnetic field components.

\subsection{2D evolution in spherical coordinates: force-free solutions}

In spherical coordinates, one of the few analytical solutions that can be used to confront numerical results is
the evolution of pure Ohmic dissipation modes (i.e., in the limit ${\cal R}_m \rightarrow 0$). If the MF 
satisfies $\curlB=\mu\vec{B}$, with constant $\mu$, 
and the magnetic diffusivity $\eta$ is constant, all components of the MF decay exponentially as 
$\propto \exp{(-t/\tau_d)}$. The family of solutions satisfying the above relation is described by radial parts involving the Bessel spherical functions and their derivatives. We test the only solution with a regular behavior at the center, the dipolar $l=1$ function of the first kind. For this solution, the MF components in the interior of a spherical domain of radius $R$ are:

\begin{eqnarray}\label{bessel}
  && B_r=\frac{B_0}{x^2} \left(\frac{\sin x}{x}-\cos x\right)\cos\theta~,\nonumber\\
  && B_\theta=\frac{B_0}{2 x^2}\left(\frac{\sin{x}}{x}-\cos x-x\sin{x}\right)\sin\theta~, \nonumber\\
  && B_\varphi=\frac{B_0 x_\star}{2 x} \left(\frac{\sin x}{x}-\cos x\right)\sin\theta~,
\end{eqnarray}
where $x=\mu r$, $x_\star=\mu R$, and $B_0$ is a normalization factor. In the limit $\mu \rightarrow 0$ we recover the
solution corresponding to a homogeneous vertical field $B_z\rightarrow B_0/3$.

Given this initial condition, we follow the evolution of the modes on several Ohmic timescales, until the field 
has been almost completely dissipated.  As boundary conditions we impose the analytical solutions for $B_\theta$ and 
$B_\varphi$ at the central cell and the external ghost cell. 
Note also that the toroidal and poloidal components of the field are completely
decoupled in the pure resistive limit, so that their evolution is totally independent \citep{pons07}. 

In Fig.~\ref{fig_bessel_decay} we compare the evolution of the numerical (crosses) and analytical (solid lines) solutions 
of $B_r$ and $B_\varphi$ at different times, for a model with $\mu R=11.5$.
They are indistinguishable in the graphic. To quantify the deviation, we
have evaluated the $L^2$-norm of the deviation from the analytic solution $\Delta \vec{B}=(\vec{B}-\vec{B}_{an})$ (normalized to the initial MF), defined as 
\begin{equation}\label{l2norm}
  L^2= \frac{\sum_{i,j} ||\Delta \vec{B}^{(i,j)}||}{\sum_{i,j} ||\vec{B}_{t=0}^{(i,j)}||}~,
\end{equation}
where the sum is performed over all grid cells $(i,j)$. The maximum of $L^2$ at any time for the same angular
resolution (60 cells) and three different radial resolutions of 88, 173, and 346 grid points is $3.7\times10^{-4}$,
$1.5\times10^{-4}$, and $1.0\times10^{-4}$ respectively.


\section{Evolution of a purely toroidal field in a neutron star crust.}\label{tor_sec}

We now consider the case of a purely toroidal MF confined to the crust $r_c<r<R$ of a realistic neutron star, with $r_c=10.8$ km and $R=11.5$ km. We refer the reader to section 4 in \cite{pons07} and section 4 in \cite{aguilera08} for details about the neutron star model and the electron density profiles, which depend only on the radial coordinate ($n_e(r)$). For simplicity, we fix a constant temperature of $10^8$ K, which gives a magnetic diffusivity in the range $\eta\in[10^{-2}-10^1]$ km$^2$/Myr. The boundary conditions we impose in this case are vanishing MF at both boundaries. According to the Hall induction equation, any initial toroidal configuration will remain purely toroidal during the evolution, but the shape and location of currents can be substantially modified. The initial MF is given by:

\begin{equation}\label{btorq}
  B_\varphi=-B_0\frac{(R-r)^2(r-r_c)^2\sin\theta\cos\theta}{r}
\end{equation}
where $B_0$ is the normalization adjusted to fix the maximum value of the toroidal field ($B_\varphi^{\rm max}$) in the numerical domain to an specific value. This corresponds to an initial maximum value of magnetic Reynolds number ${\cal R}_m^{\rm max} \approx 10$ for $B_\varphi^{\rm max}=10^{14}$ G. We use this model to highlight a few important issues concerning energy conservation, numerical viscosity, and shock formation.

\begin{figure}[t]
 \centering
 \includegraphics[width=.4\textwidth]{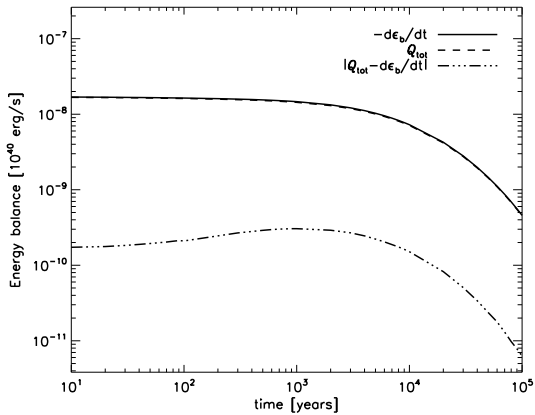}
 \caption{Volume-integrated energy balance as a function of time for a run on a 27$\times$60 grid. At each timestep loss of magnetic energy (solid line) is compensated by Joule dissipation rate (dashed), within the a small error (dot-dashed). The initial model is given by Eq.~(\ref{btorq}) with $B_\varphi^{\rm max}=10^{14}$ G.} 
 \label{fig_bcons}
\end{figure}

\subsection{Energy conservation.}

Following the definitions of Section 3.5, we check the instantaneous local conservation of energy, evaluating Eq.~(\ref{en_cons}) in the whole volume numerical domain, during a time step $\Delta t$:
\begin{equation}\label{balance}
- \sum \frac{{\bar{B^2}(t+\Delta t)}-{\bar{B^2}(t)}}{8\pi \Delta t} \Delta V = \sum \frac{4\pi \eta}{c^2} \bar{J^2} \Delta V
\end{equation}
where the sum is performed over the cells of the domain, and $\bar{B^2},\bar{J^2}$ are local averages inside each cell of volume $\Delta V$. We have used that the Poynting flux through the boundaries is zero, since the MF vanishes at both the internal boundary and the star surface. In Fig.~\ref{fig_bcons} we show the two sides of Eq.~(\ref{balance}) as a function of time (dashed and solid lines), for a model with $B_\varphi^{\rm max}=10^{14}$ G and with a coarse grid ($27\times 60$). Even for this low resolution, the error in the instantaneous energy balance (dot-dashed) is a couple of orders of magnitude less than the instantaneous magnetic energy loss (solid) and/or the Ohmic dissipation rate (dashed).

\subsection{Numerical viscosity.}

\begin{figure}[!t]
 \centering
 \includegraphics[width=.4\textwidth]{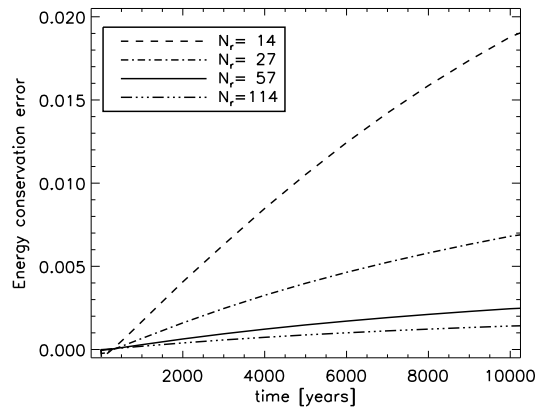}
 \includegraphics[width=.4\textwidth]{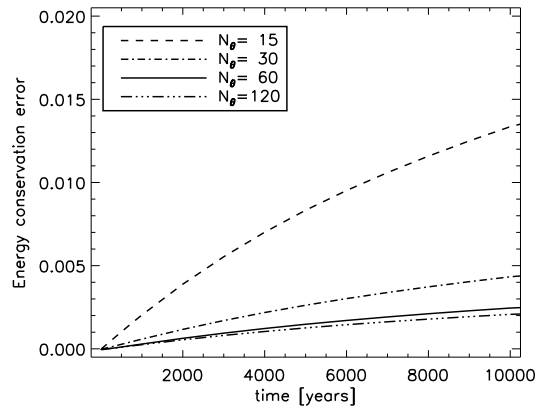}
 \caption{Accumulated relative error $\epsilon_{en}(t)$ (see Eq.~(\ref{encons_error})) in the conservation of energy during the evolution of a purely toroidal MF with $B_\varphi^{\rm max}=10^{14}$ G. We show $\epsilon_{en}(t)$ for different resolutions. Top: varying $N_r$, with fixed $N_\theta=60$; bottom: varying $N_\theta$, with fixed $N_r$.} 
 \label{fig_dissipation}
\end{figure}

It is well known that Godunov type methods provide stability in the most extreme conditions at the price of introducing numerical viscosity. Reconstruction procedures, such as the MC method we used for our code, help decrease the numerical viscosity in smooth regions of the flow and increase the spatial order of the method. However, it is important to check that any numerical viscosity introduced by the method is well below the {\it physical} resistivity, in order not to affect the results.   

In Fig.~\ref{fig_dissipation} we compare, for different resolutions, the relative error on the conservation of the total energy: 
\begin{equation}\label{encons_error}
  \epsilon_{en}(t)=1 - \frac{{\cal E}_{tot}(t)}{{\cal E}_{tot}(t=0)}
\end{equation}
where ${\cal E}_{tot}(t)= {\cal E}_b(t) + \int_0^t{\cal Q}_{tot}dt + \int_0^t{\cal S}_{tot}dt$. The top panel shows $\epsilon_{en}(t)$ with fixed $N_\theta=60$ and $N_r=14,27,57,114$ (number of points in the crust) while the bottom panel shows results for $N_r=57$ and varying angular resolution $N_\theta=15,30,60,120$. The error in the energy conservation grows nearly linear with time in the first $10^4$ yr. Assuming that this deviation is entirely due to the numerical viscosity of the method, we can estimate the viscous timescale of the numerical method by assuming that
\begin{equation}
  \epsilon_{en}(t)=1-\exp{(-t/\tau_{\rm num})}\simeq \frac{t}{\tau_{\rm num}}~,
\end{equation}
and extracting $\tau_{\rm num}$ from the results.

With a resolution of 57$\times$60 grid points (solid line in Fig.~\ref{fig_dissipation}), the error on energy conservation is of 0.25 $\%$ at $10^4$ yr, which implies 
$\tau_{\rm num} \sim 4 $ Myr. Therefore we can estimate the order of magnitude of the numerical resistivity as
\begin{equation}
{\eta_{\rm num}} =\frac{\Delta r^2}{\tau_{\rm num}}~,
\end{equation}
where $\Delta r \sim (r_\star - r_{core})/N_r$ is the average radial size of a cell. For $N_r=57$, we obtain $\eta_{\rm num}\sim 10^{-4}$ km$^2$/Myr. Comparing this value with the typical physical resistivity of the crust at this temperature, $\eta\sim 10^{-2}-10^1$ km$^2$/Myr, we can assess that, even with a coarse grid of 57$\times$60 grid points, the numerical viscosity is at least two orders of magnitude less than the physical resistivity.

We have also studied the effect on the numerical viscosity of the Courant prefactor $k_c$, defined in Eq.~(\ref{timestep}). We find that varying $k_c$ does not affect significantly the numerical dissipation. This effect is always much smaller than the dependence on the spatial resolution, as long as the evolution is stable (typically $k_c<0.1$).

\subsection{Creation of current sheets.}

\begin{figure}[t]
 \centering
 \includegraphics[width=.22\textwidth]{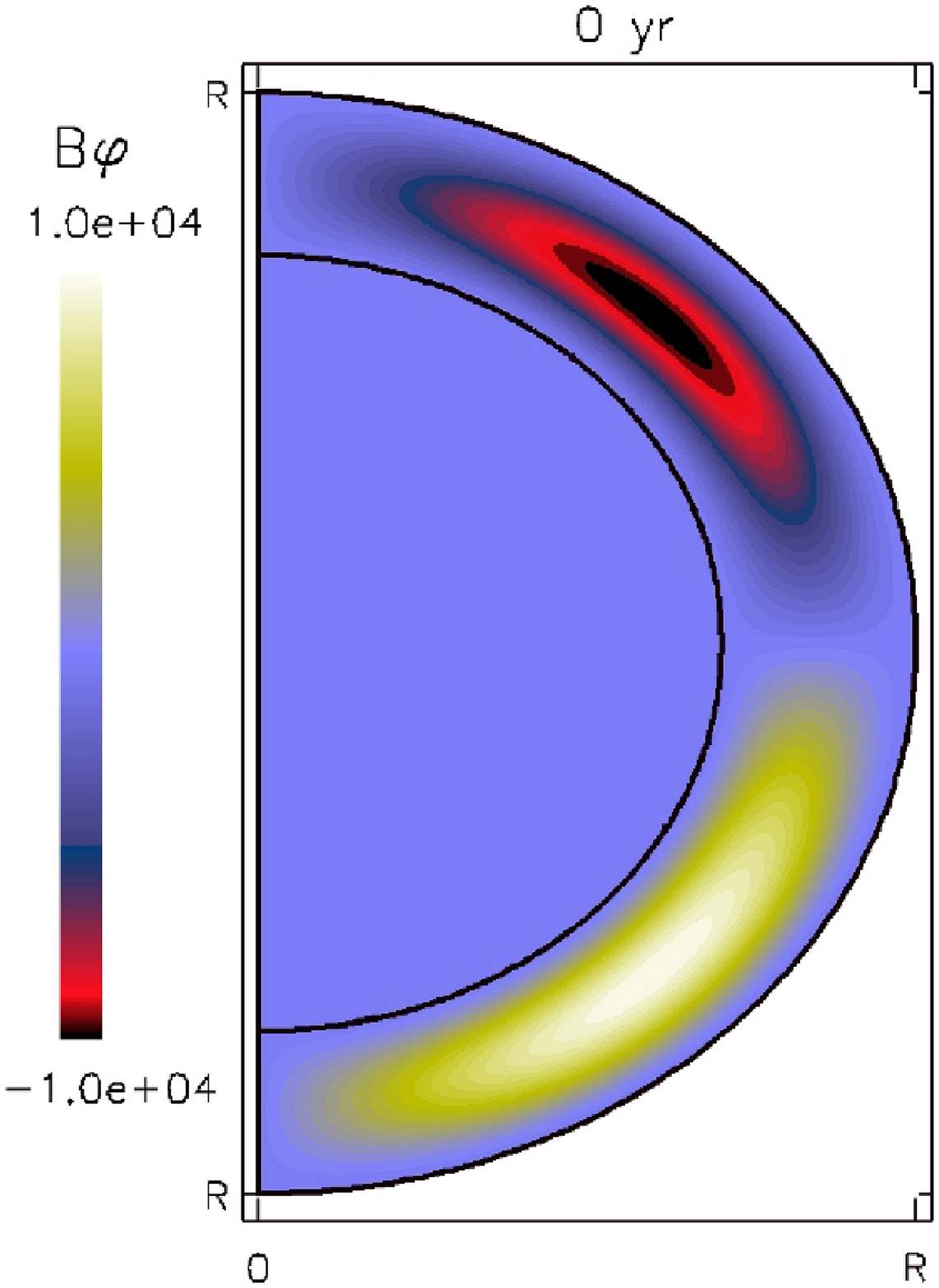}
 \includegraphics[width=.22\textwidth]{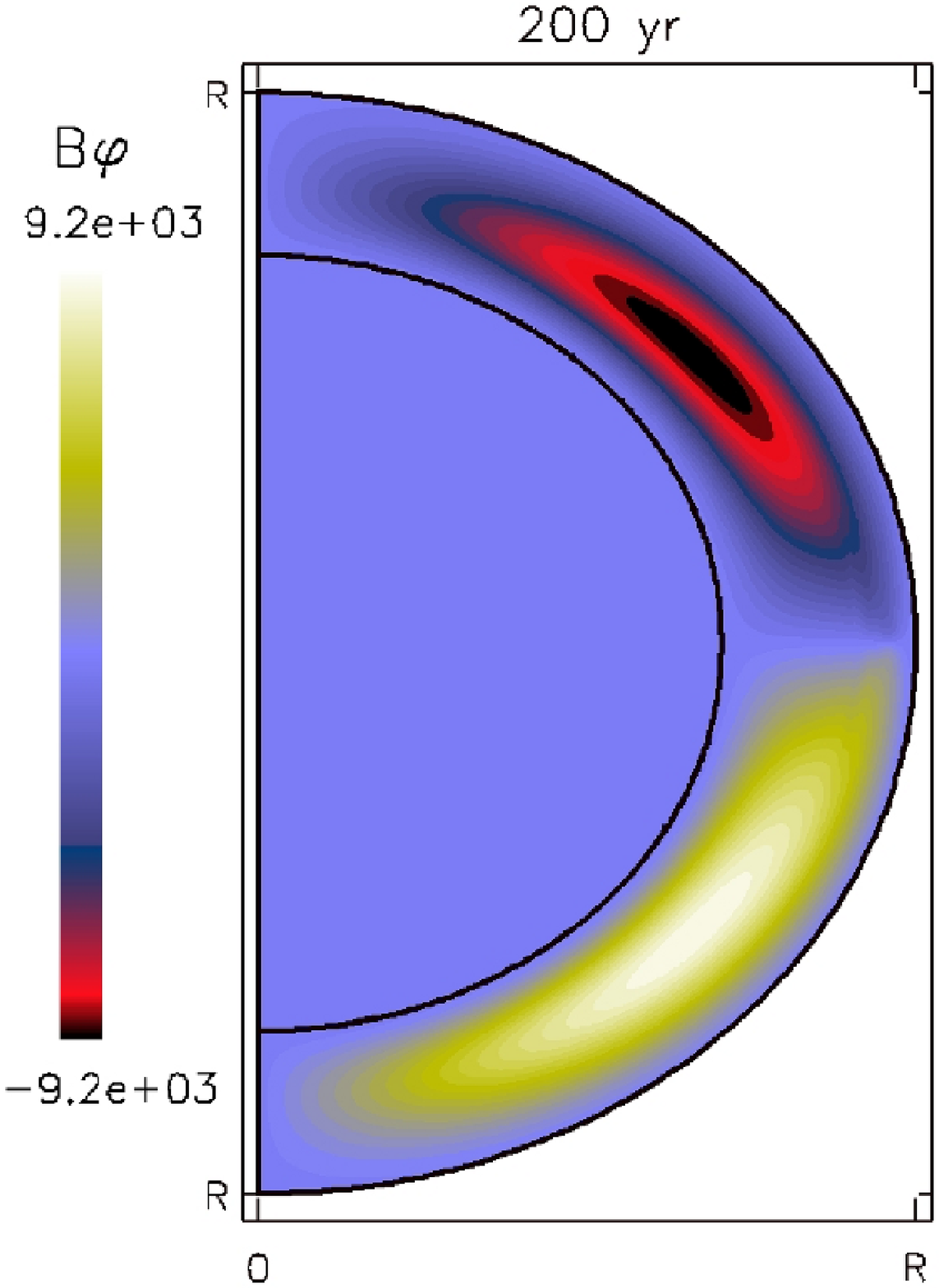}
 \includegraphics[width=.22\textwidth]{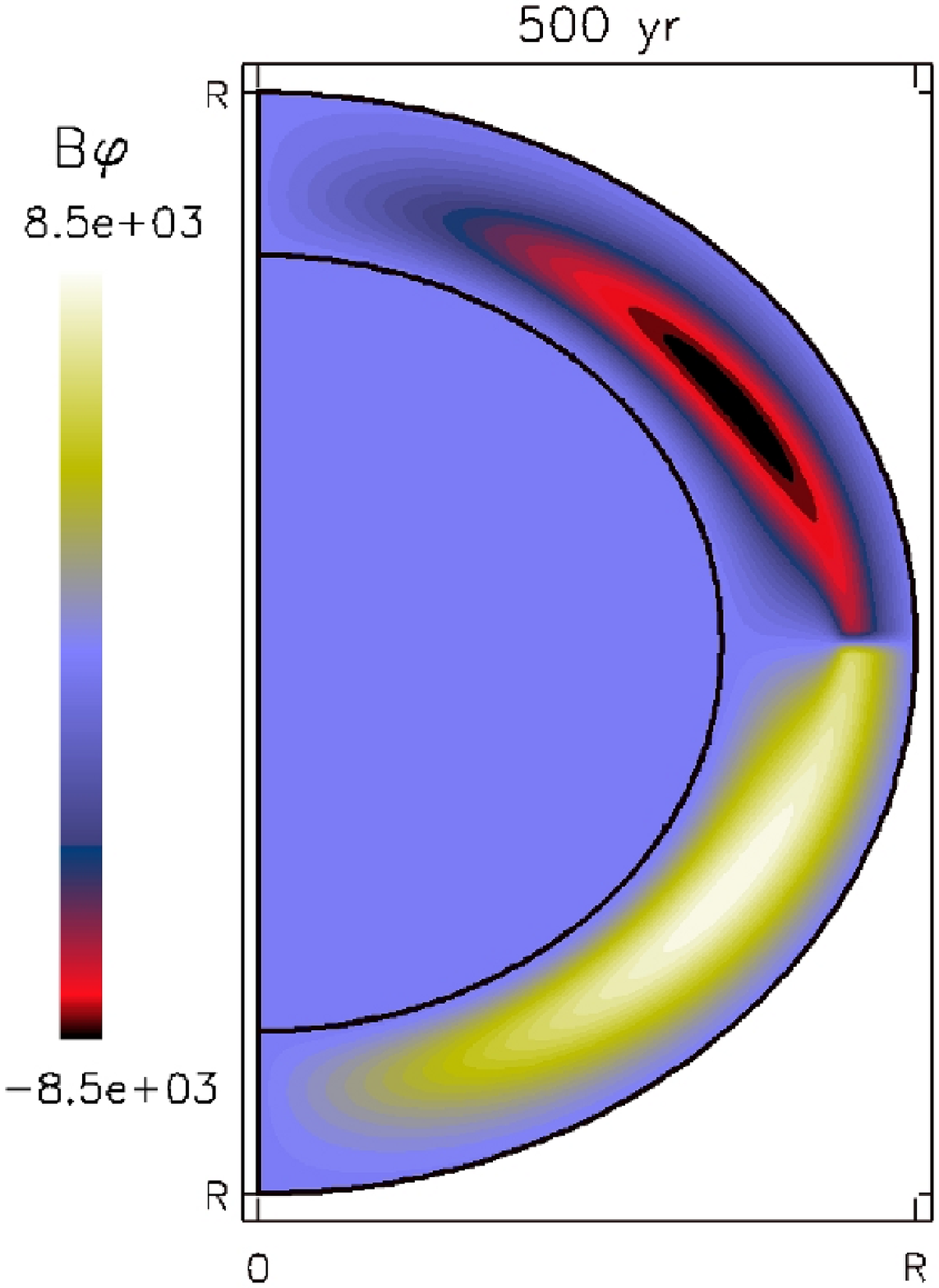}
 \includegraphics[width=.22\textwidth]{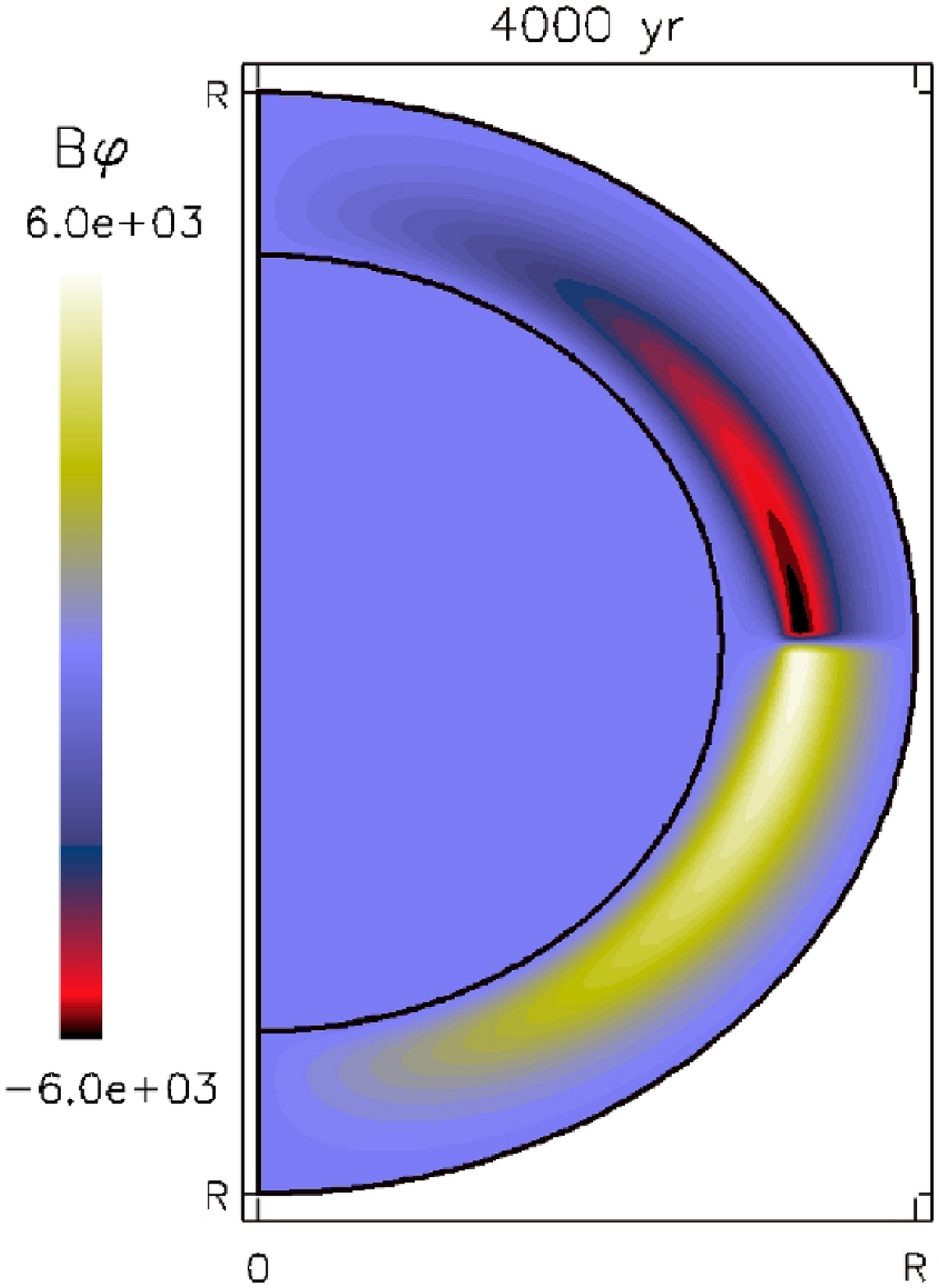}
\caption{Evolution of a quadrupolar toroidal magnetic field confined into the neutron star crust. We show snapshots at $t=0, 200, 500,$ and 4000 yr. The color scale indicates the toroidal field strength, in units of $10^{12}$ G. In the figure, the thickness of the crust has been stretched by a factor of 4 for clarity.} 
 \label{fig_bphi}
\end{figure}

\begin{figure}[t!]
 \centering
 \includegraphics[width=.45\textwidth]{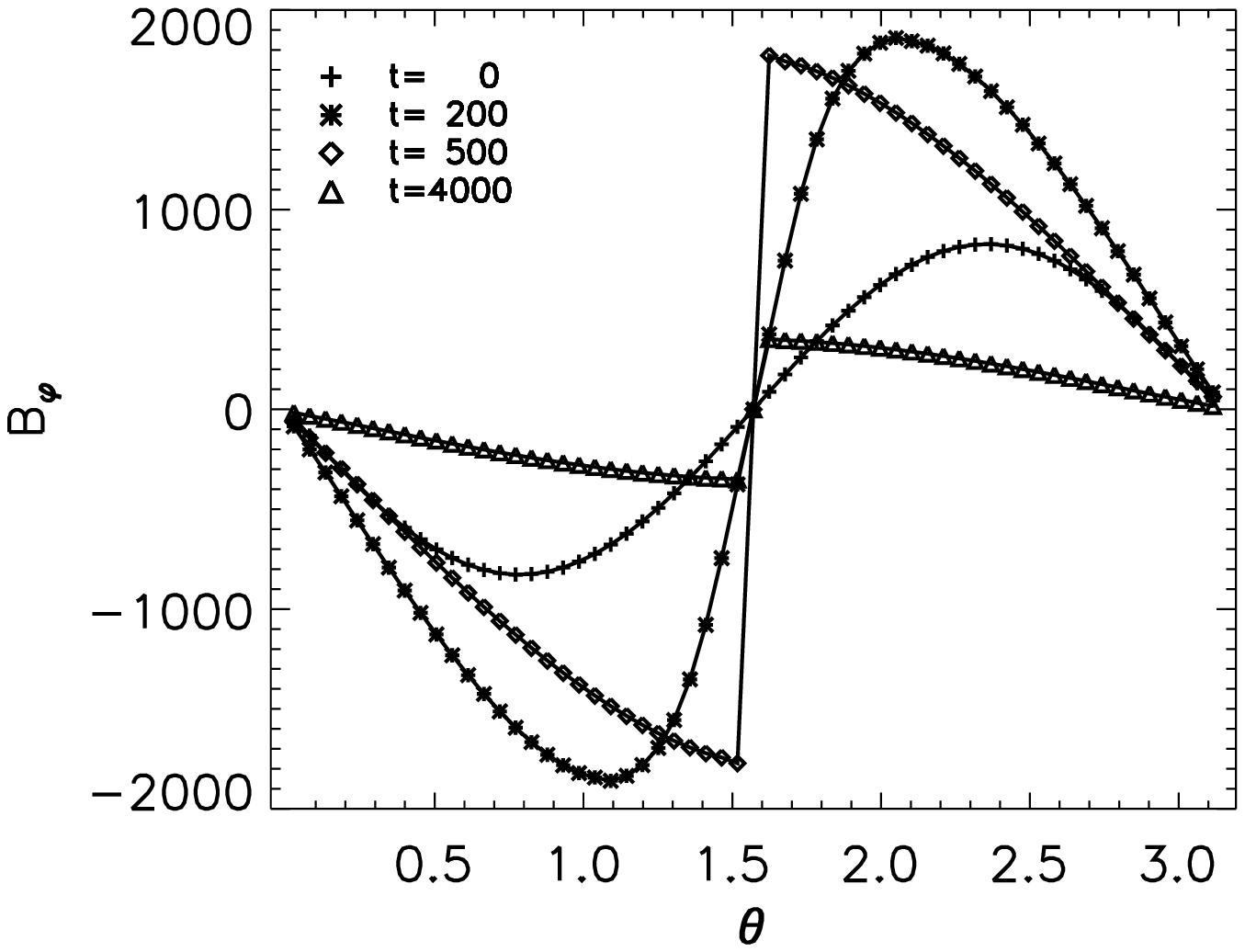}
 \includegraphics[width=.45\textwidth]{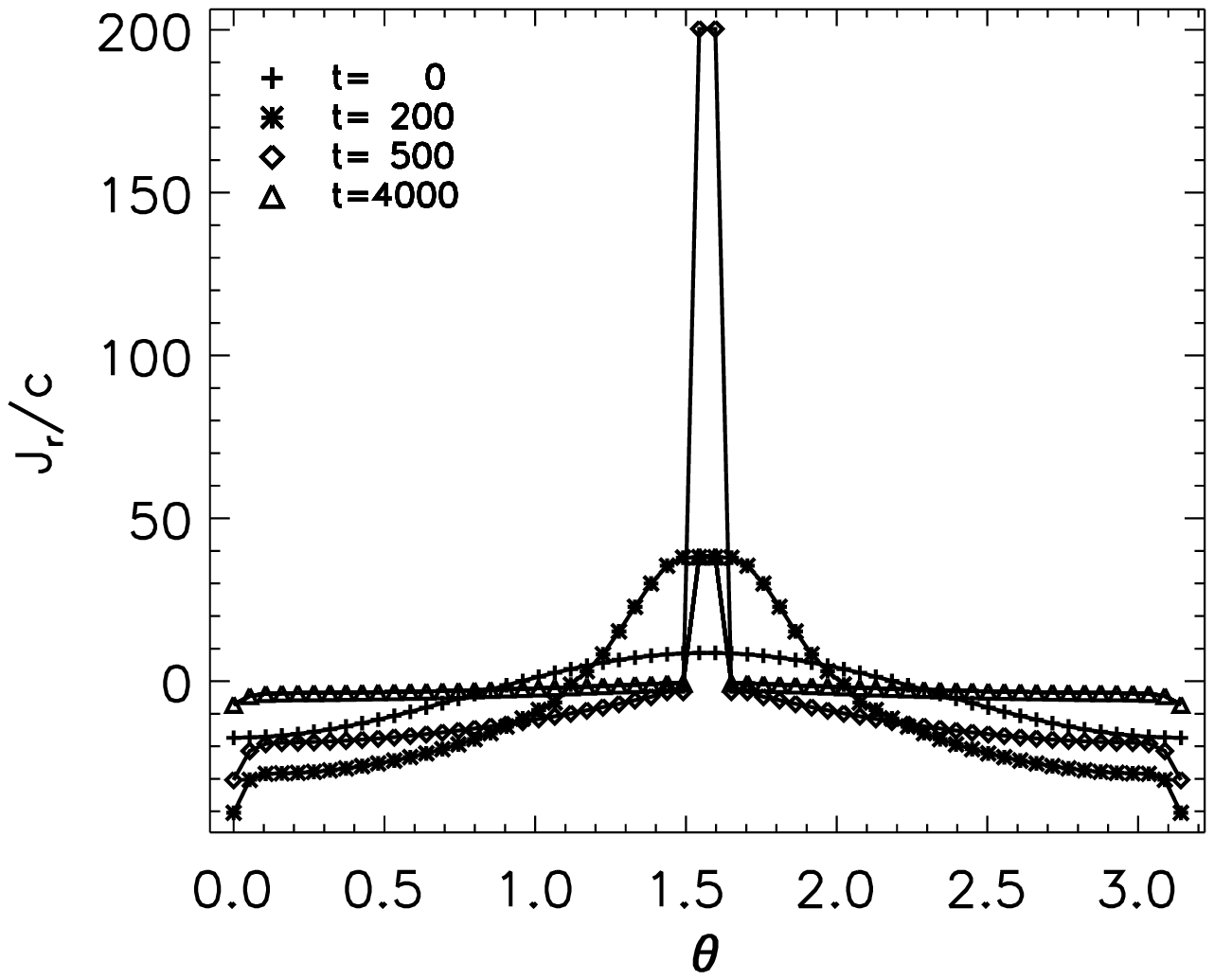}
\caption{Angular profiles of the same model as Fig.~\ref{fig_bphi}, just below the star surface, of $B_\varphi$ (in units of $10^{12}$ G) and $J_r/c$ (in units of $10^{12}$ G/km), at four different times $t=0, 200, 500,$ and 4000 yr. The angular resolution used is $N_\theta=60$.} 
\label{fig_shock_up}
\end{figure}

\begin{figure}[t]
 \centering
 \includegraphics[width=8cm]{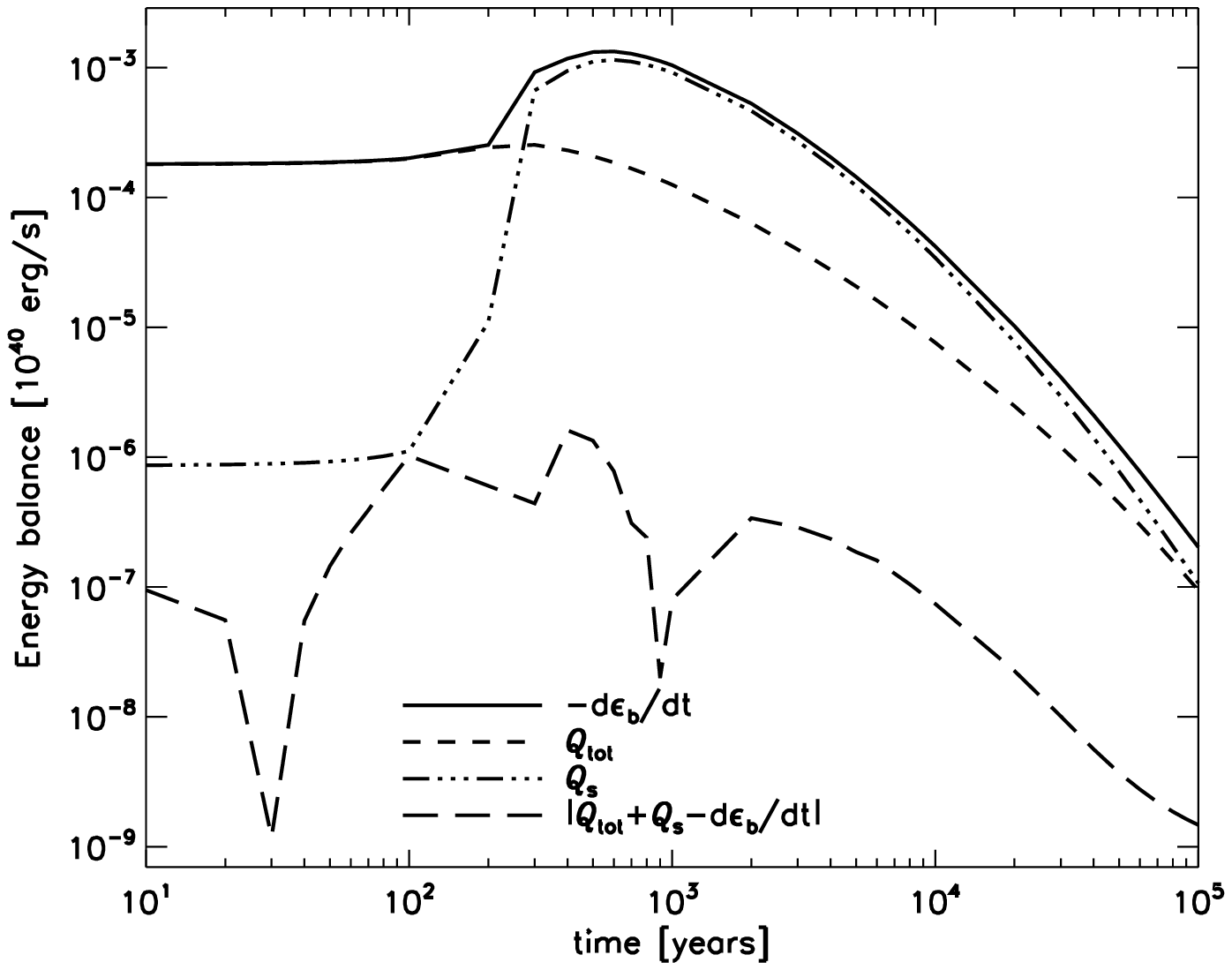}
 \includegraphics[width=8cm]{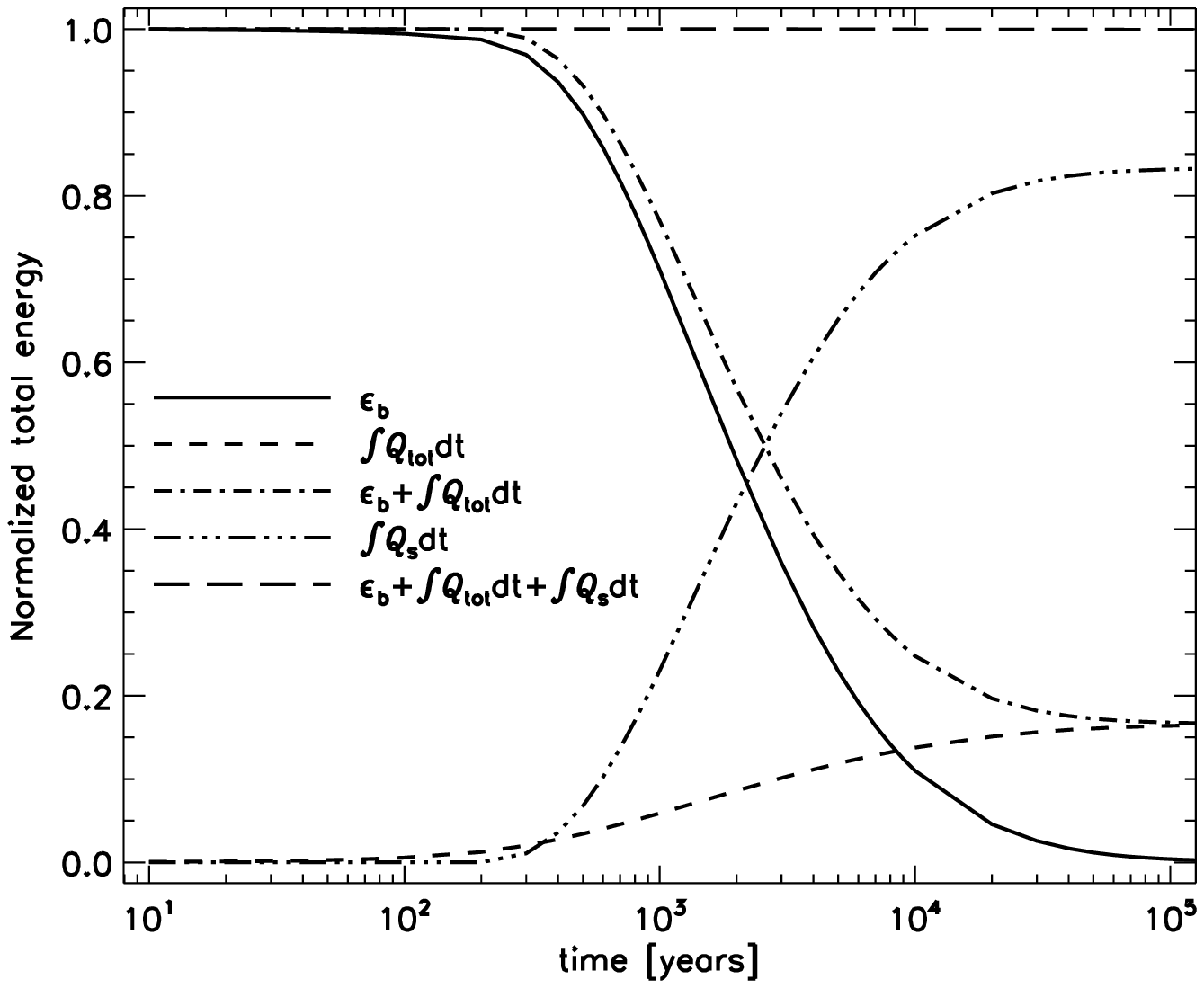}
\caption{Instantaneous (top) and time-integrated (bottom) energy balance for a purely toroidal MF, with $114\times200$ crustal points. The variation of the magnetic energy (solid line), Joule dissipation rate (dashes) and dissipation rate at the shock (triple-dot-dash) are shown together with the total energy balance (long dashes). The bottom panel also shows the sum (dash-dot) of the magnetic energy and energy dissipated by Joule effect.}
 \label{fig_en_shock}
\end{figure}

The evolution equation for a purely toroidal MF, neglecting resistivity, can be cast into the following form:
\begin{equation}\label{hall_burger}
 \frac{\partial B_\varphi}{\partial t} + \lambda_r n_e\frac{\partial}{\partial r}\left(\frac{B_\varphi^2}{2 n_e}\right) + \lambda_\theta\frac{1}{r}\frac{\partial}{\partial \theta}\left(\frac{B_\varphi^2}{2}\right) = 0
\end{equation}
where
\begin{eqnarray}\label{lambda_burgers}
 && \lambda_r=-\frac{c}{2 \pi e n_e}\frac{\cos\theta}{r\sin\theta} \nonumber\\
 && \lambda_\theta=-\frac{r^2 c}{4 \pi e}\frac{d}{dr}\left(\frac{1}{n_e r^2}\right)~.
\end{eqnarray}
This form explicitly shows the Burgers-like character of the equation. In particular, a realistic profile of $n_e(r)$ has a steep gradient, which may result in the the fast formation of a current sheet. The location where it forms will depend on the initial geometry. This hidden character will thus become evident when ${\cal R}_m \gg 1$.

In order to test this limit in spherical coordinates, we study the same model as in the previous section but with a larger field, taking $B_\varphi^{\rm max}=10^{16}$ G (implying ${\cal R}_m^{\rm max} \approx 1000$). Fig.~\ref{fig_bphi} shows four snapshots of the evolution of such configuration, where the dominant role of the Hall drift
term leads to the fast formation of a current sheet in the equatorial plane. In Fig.~\ref{fig_shock_up} we show a sequence of angular profiles of $B_\varphi$ (top) and $J_r/c$ (bottom) close to the star surface. Our numerical code is able to follow the ``shock'' formation and captures the thin current sheet with only two grid points.

In Fig.~\ref{fig_en_shock} we show the contribution of the different terms in the energy balance equation as a function of time. In the bottom panel, the sum of the magnetic energy (solid) and the energy dissipated by the Joule effect (short dashes) is shown by a dash-dot line. The sum of these two terms is constant and equal to the initial magnetic energy for smooth solutions, as shown in the previous subsection. We can observe that now this is very accurately satisfied until $\approx 200$ yr, coinciding with the formation of the shock. After this point, the local energy balance calculated as above seems to be violated. About $90\%$ of the magnetic energy is dissipated in $10^4$ yr, but the energy dissipated by the Joule effect is only $13\%$ of the initial magnetic energy, yielding to an apparent loss of total energy of $\sim 77\%$. This seemingly incorrect result has a physical explanation: when a current sheet forms, there is energy dissipated at the discontinuity not accounted for by the standard Joule dissipation rate (right-hand side of Eq.~(\ref{balance})). We find the same result with $N_\theta=60,120,$ or 200. Our numerical scheme can capture the effect of an extremely thin current sheet (middle panel of Fig.~\ref{fig_shock_up}) even if we are dealing with a low resolution. However, we unavoidably underestimate the dissipation at the discontinuity, which affects the total energy balance, although the implied volume is small.

A rigorous mathematical study of the energy dissipation rate at the shock in one-dimensional Burgers flows, governed by $\partial_t u + u\partial_x u=\nu \partial_{xx}u$, shows that, in the inviscid limit $\nu=0$, the total energy dissipation rate goes as $2[u]^3/3$, where $[u]$ denotes the half-jump in the variable $u$ across the shock \citep{tran10}.  By analogy with the results obtained for one-dimensional Burgers flows, we propose an estimate of the energy dissipated at the shock across a surface $\Sigma_\alpha$:
\begin{equation}
 {\cal Q}_s=-\frac{1}{6\pi}\lambda_r[B_\varphi]^3\Sigma_r~,
\end{equation}
if the discontinuity is in the radial direction, or
\begin{equation}\label{qs_ang}
 {\cal Q}_s=-\frac{1}{6\pi}\lambda_\theta[B_\varphi]^3\Sigma_\theta~,
\end{equation}
if it is in the angular direction. This correction is accounting for the magnetic flux lost across the surface of the shock. The coefficients $\lambda_r$, $\lambda_\theta$ are defined in Eqs.~(\ref{lambda_burgers}). Note that the correction has to be applied only at the interfaces where $\lambda_\alpha[B_\varphi]$ is negative, which is the necessary condition for having a shock. In Fig.~\ref{fig_en_shock} we show the the effect of the above correction in the energy balance when a discontinuity in the toroidal field is detected. Initially, when the MF is smooth, the extra term (triple-dot-dash) plays no role. It should vanish, but discretization errors result in a $\lesssim 1\%$ correction. However, when the current sheet is formed, it becomes the dominant contribution to explain the loss of magnetic energy. Applying this correction, the total energy balance is very well satisfied. After $10^5$ yr, when the MF has been almost completely dissipated, the error in the total energy conservation is only $\epsilon_{en}\sim 7 \times 10^{-4}$.

Therefore, this estimation of the dissipated energy at the shock turns out to be an excellent approximation.
Note that in this particular case (a purely toroidal field forming a current sheet exactly at the equator, with the discontinuity
in the angular direction) the generalization of the
planar exact result to spherical coordinates is a very good approximation, since each radial layer is governed by
an independent one-dimensional Burgers-like equation in the vicinity of the equator. In more general cases (asymmetric with
respect to the equator, or including poloidal components) this approximation 
must be taken only as an estimate of the energy dissipation rate, within a factor of two.
This term is of great importance for realistic magneto-thermal evolution models, because it strongly enhances the local deposition of heat.

\section{A general example}\label{all_sec}

\begin{figure}[t!]
 \centering
 \includegraphics[width=.23\textwidth]{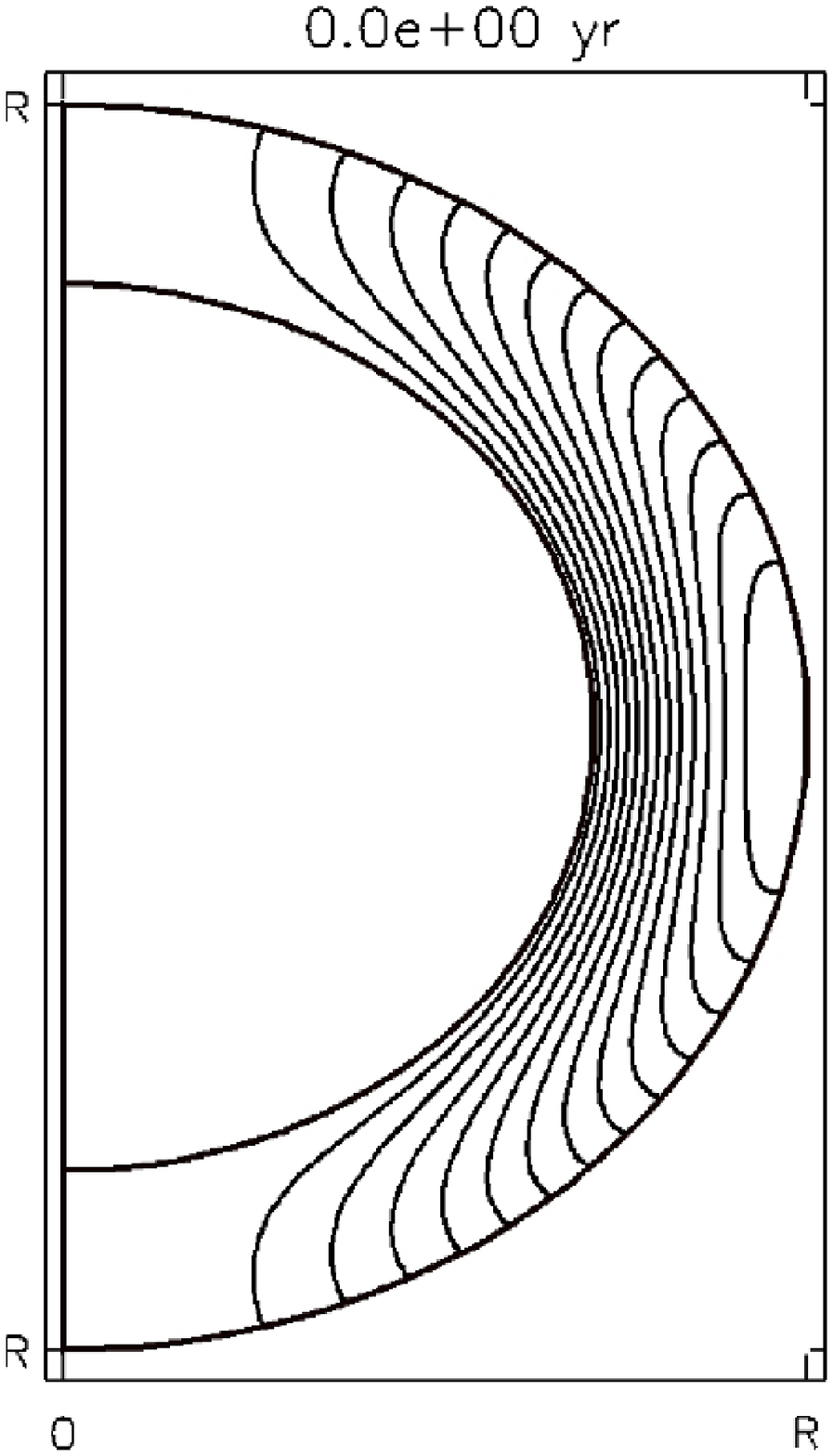}
 \includegraphics[width=.23\textwidth]{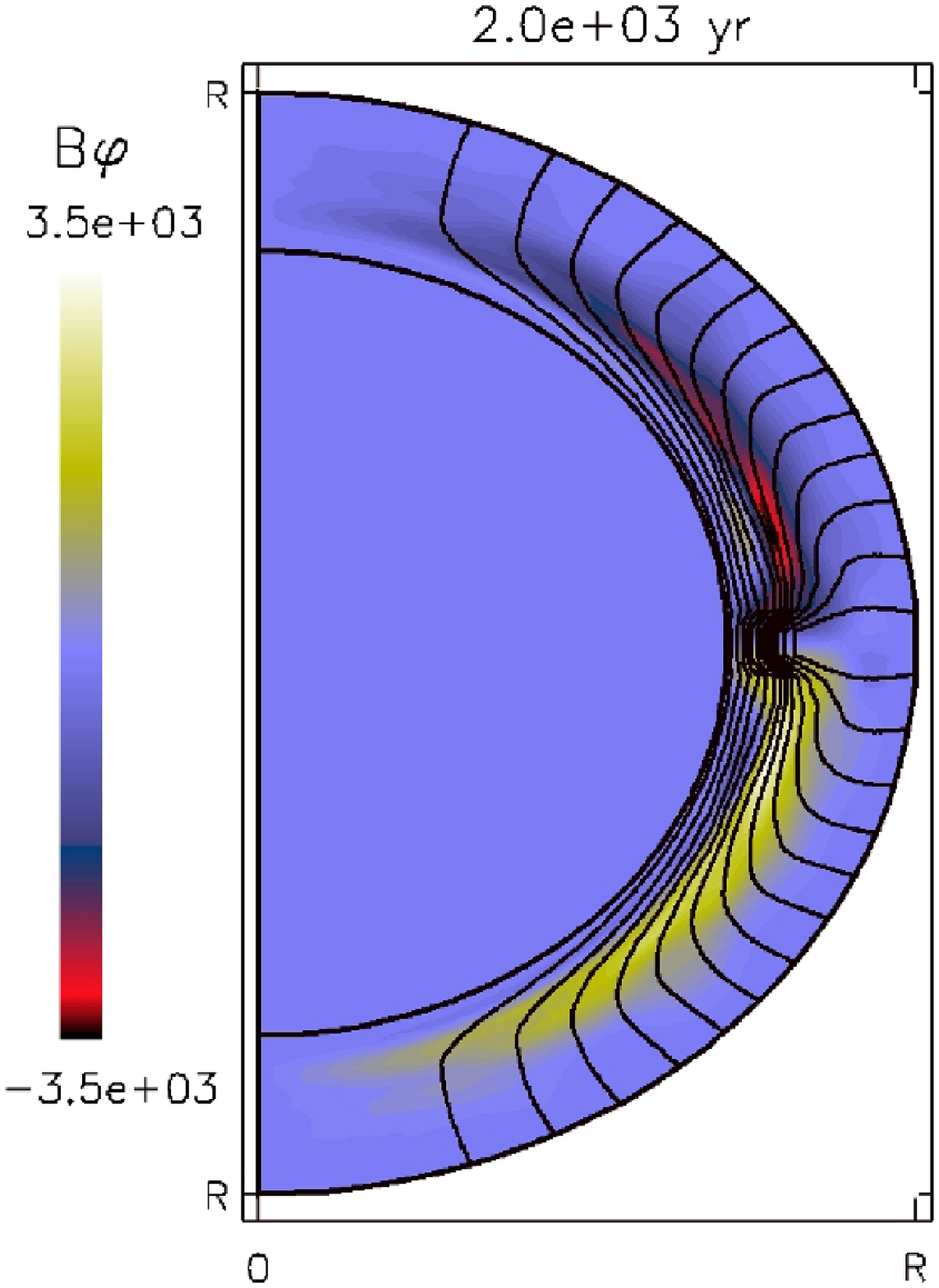}
 \includegraphics[width=.23\textwidth]{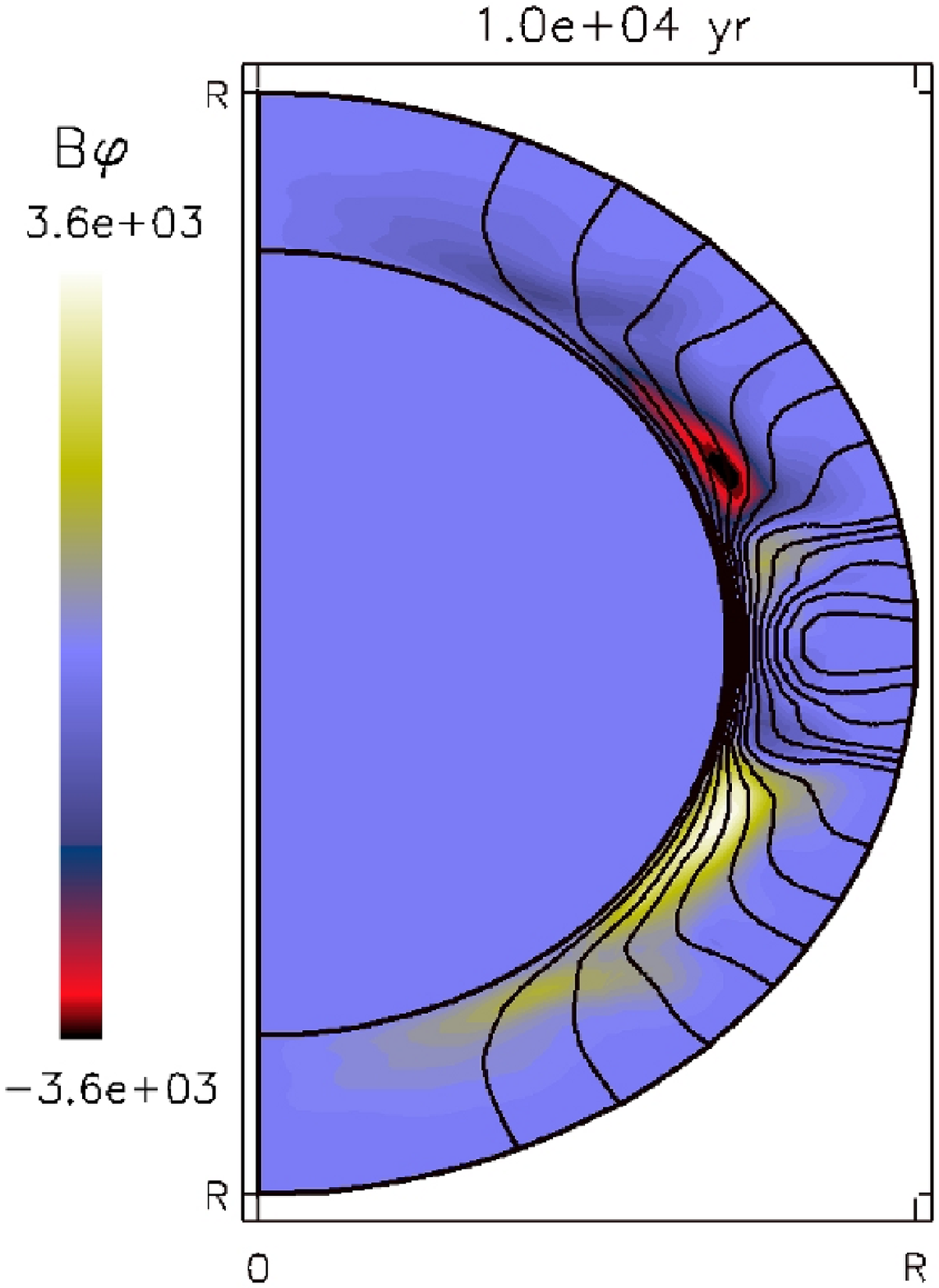}
 \includegraphics[width=.23\textwidth]{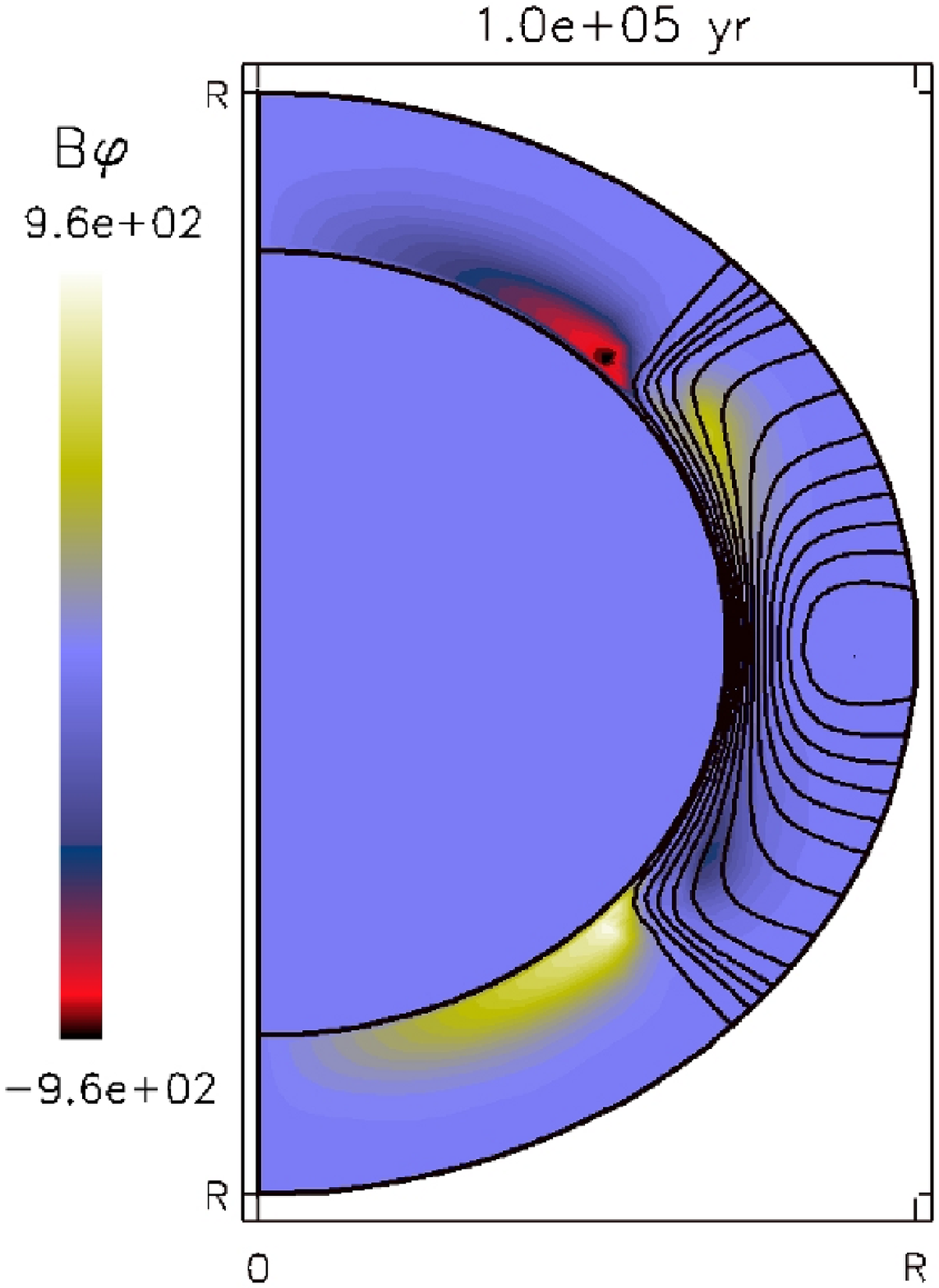}
\caption{Evolution of an initially purely poloidal magnetic field. The figure shows snapshots at $t=0, 2, 10,$ and 100 kyr. Solid lines show the poloidal field lines and the color scale indicates the toroidal field strength, in units of $10^{12}$ G. In the figure, the thickness of the crust has been stretched by a factor of 4 for clarity.} 

 \label{fig_mix}
\end{figure}

The code can handle any MF geometry, confined to the crust or extended to the center, and has been tested in all the limiting cases. To conclude our study, we show results for a general case consisting in a crustal confined magnetic field with both toroidal and poloidal components.

We apply vacuum boundary conditions by setting $B_\varphi =0$ for $r>R$ and matching the poloidal field with a multipolar vacuum solution. The vacuum solution is imposed as a von Neumann boundary condition problem for a potential magnetic field, expressed as $\vec{B}=\vec{\nabla}\Psi$, where $\Psi$ is the magnetic scalar potential. Using the Green's representation formula \citep{jacksonbook} with $\Psi$ and the choice of Green's function $G(\vec{r},\vec{r}')=|\vec{r}-\vec{r}'|^{-1}$, the relation between $B_r$ and $B_\theta$ at surface $r=R$ can be expressed as (M. Reinhardt, priv. comm.):

$$ 4\pi \int_0^\theta B_\theta(R,\theta'){\rm d}\theta' + \int_0^\pi B_\theta(R,\theta')\int_{\theta'}^\pi f(\theta,\theta''){\rm d}\theta''{\rm d}\theta' = $$
\begin{equation}
 = - 2\int_0^\pi B_r(R,\theta') f(\theta,\theta'){\rm d}\theta' \label{bvacuum_green}~,
\end{equation}
where
$$ f(\theta,\theta')=\int_0^{\pi/2}\frac{\sqrt{8}\sin\theta' \quad{\rm d}\phi'}{\sqrt{1-\cos(\theta-\theta')+2\sin\theta\sin\theta'\sin^2\phi'}}~.$$
Alternatively, the vacuum solution can be expressed in terms of Legendre polynomials ($P_l$) as follows:
\begin{eqnarray}
  && B_r      =   \sum_l b_l (l+1)P_l(\cos\theta) \left(\frac{R}{r}\right)^{-(l+2)}~, \nonumber \\
  && B_\theta = - \sum_l b_l \frac{dP_l(\cos\theta)}{d\theta} \left(\frac{R}{r}\right)^{-(l+2)}~, \label{bvacuum}
\end{eqnarray}
where $b_l$ are the weights of the multipoles, obtained from the Legendre decomposition of $B_r(R,\theta)$:
\begin{equation}
 b_l  = \frac{2l+1}{2(l+1)} \int_{0}^{\pi} B_r(R,\theta) P_l(\cos\theta) d\theta  ~,
\end{equation}
At each time step, we cast $B_r(R,\theta)$ into Eq.~(\ref{bvacuum_green}) or Eq.~(\ref{bvacuum}) to reconstruct the value of $B_\theta$ in the first external ghost cell. We have verified that, for a given $B_r(R,\theta)$, both methods provide the same function $B_\theta(R,\theta)$, reproducing correctly the analytical cases (e.g., a dipole). In any case, the vacuum boundary condition is equivalent to avoid current escaping (entering) from (into) the star. The non-vanishing Poynting flux across the outer boundary allows the (small) interchange of magnetic energy with the external field. For the inner boundary, we set superconducting boundary conditions (i.e. $B_r=0$ and vanishing tangential electric field $E_\theta=E_\varphi=0$) at the crust/core interface. As a consequence, the Poynting flux is zero and no energy is allowed to flow into/from the superconductive core.

Our initial model consists of a purely dipolar poloidal component built from the following potential vector
\begin{equation}\label{aphi}
 A_\varphi(r,\theta)=-f(r)\sin\theta~,
\end{equation}
where $f(r)$ is a radial function smoothly matching with the vacuum solution at $r=R$ and vanishing radial component at $r=r_c$ (see \S 2.1 of \cite{aguilera08} for details). We normalize the MF intensity at the pole to $B_p= 10^{15}$ G and fix the temperature to $T=10^8$ K during the evolution. This implies initial maximum values of ${\cal R}_m\sim 150$.

Fig.~\ref{fig_mix} shows the initial configuration (top left panel) and three snapshots of the evolution of the described model. Because of the Hall term, the initially poloidal field immediately develops a toroidal component of opposed parity (mainly quadrupolar) on a timescale of about $~2\times 10^3$ yr. 
Note that in this case the MF components must remain (anti)symmetric with respect to the equator.
When the toroidal field grows to values similar to the poloidal field strength, the Hall term also leads to the displacement of MF lines to the equator and the formation of a discontinuity in $\theta$-direction, as discussed in previous sections, but now in both $B_r$ and $B_\varphi$. After that, whistler waves are launched towards the poles and some sort of oscillatory mode appears, with transfer of energy between the poloidal and toroidal components. The structures are well resolved even with a coarse grid of $60\times 27$ grid, and the discontinuities survive until the field is dissipated and the relative importance of the Hall term decreases, on timescales of $\sim 10^5$ yr.

\begin{figure}[t!]
 \centering
 \includegraphics[width=.4\textwidth]{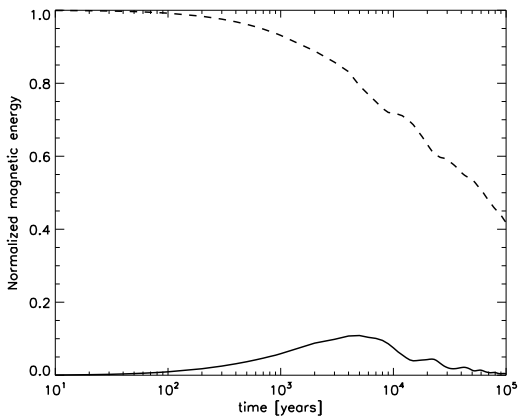}
\caption{Magnetic energy stored in the toroidal (solid line) and poloidal field (dashed) during the evolution.} 
 \label{fig_transfer}
\end{figure}

Fig.~\ref{fig_transfer} shows the magnetic energy stored in the toroidal and poloidal component, normalized to the initial magnetic energy. During the first $5\times 10^3$ yr, the toroidal field gains energy from the poloidal field, until it reaches a $10\%$ of the initial MF energy. Then, a quasi-periodic transfer of energy between poloidal and toroidal field is observed. The toroidal field strength exceeds locally the values of the poloidal field. The relatively low energy of the toroidal component is due to the small volume of the regions where its intensity is large. Both components acquire a marked multipolar behavior, with discontinuities located not only in the equator. The particular details about how much energy can be interchanged between poloidal and toroidal components strongly depend on both the geometry of the initial configuration and the MF strength. For weak fields (low ${\cal R}_m$) when the resistive term dominates, the two components are nearly decoupled and the transfer of energy between them is negligible. In situations dominated by the Hall terms, it can be significant. We defer a deeper discussion about the astrophysical implications for future work.

\section{Conclusions}\label{sec_conclusions}

We have described a new code for the EMHD evolution of magnetic field of Neutron Stars. The code implements algorithms based on higher order Godunov methods with a discretization to evolve area averages of the face-centered components of the magnetic field (magnetic fluxes). This combination results in a divergence-preserving scheme able to deal with discontinuities and follow the formation of current sheets.

The primary motivation for developing this code has been the need to follow the evolution of the internal magnetic field in neutron stars, including the Hall term, with an arbitrarily high magnetic Reynolds number. We also need to resolve magnetic field structures over a wide range of length scales and the formation of current sheets in the astrophysical applications of interest. In this paper, we have focused on the detailed implementation of the methods into a functioning computer code with emphasis in the importance of using the staggered grid and the conservative formulation, and the need to consider the Burgers-like character of the equations. An extensive series of test problems in 2D, and for both Cartesian and spherical coordinates have been presented. These tests, that include the limiting cases of purely resistive evolution and purely Hall evolution (vanishing electrical resistivity), should be useful to others developing and testing codes for similar astrophysical scenarios. We have shown the importance of dissipation in the location of current sheets, which is usually overlooked, or incorrectly described. The analysis in \cite{tran10} shows that, using spectral methods, the number of modes needed to solve the problem scales as ${\cal R}_m$, which would require hundreds or even thousands of modes to tackle some typical problems with ``magnetar conditions''. However, we have shown in section 4.1.3 that using upwind difference schemes to solve the equations in conservative form can satisfactorily deal with the purely inviscid limit with a coarse grid. We have also proposed a simple generalization of the exact solution found for the 1D Burgers equation to estimate the energy dissipation rate at the location of a discontinuity, wherever it forms. This may have implications in the evolution of neutron stars that will be discussed in future works.

We have shown that the code is stable and can follow to late times the evolution of the internal magnetic field in neutron stars with very low temperature, which had never been possible before. In future works, we will move beyond the developmental phase and we plan to use the code for a variety of applications, including the evolution of magnetic field in the crust of magnetars, the study of field submergence by accretion of matter, and magnetic flux expulsion form the core. We expect that our code will become a fundamental tool for astrophysical applications concerning neutron star magneto-thermal evolution. We are also confident that the detailed description of the algorithms provided in this paper will be useful to other researchers in the field for solving many problems in neutron stars.

\section*{Acknowledgements}
This work was partly supported by the Spanish grant AYA 2010-21097-C03-02 and CompStar, a Research Networking Program of the European Science Foundation. D.~Vigan\`o is supported by a fellowship from the \textit{Prometeo} program for research groups of excellence of the Generalitat Valenciana (Prometeo/2009/103). We thank U.~Geppert and M.~Rheinhardt for the helpful comments and contributions.

\bibliography{mt_code}

\begin{thebibliography}{21}
\expandafter\ifx\csname natexlab\endcsname\relax\def\natexlab#1{#1}\fi
\providecommand{\bibinfo}[2]{#2}
\ifx\xfnm\relax \def\xfnm[#1]{\unskip,\space#1}\fi
\bibitem[{{Aguilera} et~al.(2008){Aguilera}, {Pons} and
  {Miralles}}]{aguilera08}
\bibinfo{author}{D.N. {Aguilera}}, \bibinfo{author}{J.A. {Pons}},
  \bibinfo{author}{J.A. {Miralles}}, \bibinfo{journal}{\aap}
  \bibinfo{volume}{486} (\bibinfo{year}{2008}) \bibinfo{pages}{255--271}.
\bibitem[{{Ant{\'o}n} et~al.(2006){Ant{\'o}n}, {Zanotti}, {Miralles},
  {Mart{\'{\i}}}, {Ib{\'a}{\~n}ez}, {Font} and {Pons}}]{anton06}
\bibinfo{author}{L.~{Ant{\'o}n}}, \bibinfo{author}{O.~{Zanotti}},
  \bibinfo{author}{J.A. {Miralles}}, \bibinfo{author}{J.M. {Mart{\'{\i}}}},
  \bibinfo{author}{J.M. {Ib{\'a}{\~n}ez}}, \bibinfo{author}{J.A. {Font}},
  \bibinfo{author}{J.A. {Pons}}, \bibinfo{journal}{\apj} \bibinfo{volume}{637}
  (\bibinfo{year}{2006}) \bibinfo{pages}{296--312}.
\bibitem[{{Cerd{\'a}-Dur{\'a}n} et~al.(2008){Cerd{\'a}-Dur{\'a}n}, {Font},
  {Ant{\'o}n} and {M{\"u}ller}}]{cerdaduran08}
\bibinfo{author}{P.~{Cerd{\'a}-Dur{\'a}n}}, \bibinfo{author}{J.A. {Font}},
  \bibinfo{author}{L.~{Ant{\'o}n}}, \bibinfo{author}{E.~{M{\"u}ller}},
  \bibinfo{journal}{\aap} \bibinfo{volume}{492} (\bibinfo{year}{2008})
  \bibinfo{pages}{937--953}.
\bibitem[{{Giacomazzo} and {Rezzolla}(2007)}]{giacomazzo07}
\bibinfo{author}{B.~{Giacomazzo}}, \bibinfo{author}{L.~{Rezzolla}},
  \bibinfo{journal}{Classical and Quantum Gravity} \bibinfo{volume}{24}
  (\bibinfo{year}{2007}) \bibinfo{pages}{235}.
\bibitem[{{Goldreich} and {Reisenegger}(1992)}]{goldreich92}
\bibinfo{author}{P.~{Goldreich}}, \bibinfo{author}{A.~{Reisenegger}},
  \bibinfo{journal}{\apj} \bibinfo{volume}{395} (\bibinfo{year}{1992})
  \bibinfo{pages}{250--258}.
\bibitem[{{Hollerbach} and {R{\"u}diger}(2002)}]{hollerbach02}
\bibinfo{author}{R.~{Hollerbach}}, \bibinfo{author}{G.~{R{\"u}diger}},
  \bibinfo{journal}{\mnras} \bibinfo{volume}{337} (\bibinfo{year}{2002})
  \bibinfo{pages}{216--224}.
\bibitem[{{Hoyos} et~al.(2008){Hoyos}, {Reisenegger} and {Valdivia}}]{hoyos08}
\bibinfo{author}{J.~{Hoyos}}, \bibinfo{author}{A.~{Reisenegger}},
  \bibinfo{author}{J.A. {Valdivia}}, \bibinfo{journal}{\aap}
  \bibinfo{volume}{487} (\bibinfo{year}{2008}) \bibinfo{pages}{789--803}.
\bibitem[{{Huba}(2003)}]{huba03}
\bibinfo{author}{J.D. {Huba}}, in: \bibinfo{editor}{{J.~B{\"u}chner, C.~Dum, \&
  M.~Scholer}} (Ed.), \bibinfo{booktitle}{Space Plasma Simulation}, volume
  \bibinfo{volume}{615} of \textit{\bibinfo{series}{Lecture Notes in Physics,
  2003}}, pp. \bibinfo{pages}{166--192}.
\bibitem[{Jackson(1991)}]{jacksonbook}
\bibinfo{author}{J.D. Jackson}, \bibinfo{title}{Classical Electrodynamics},
  \bibinfo{publisher}{John Wiley \& Sons, Inc.}, \bibinfo{address}{New Jersey,
  USA}, \bibinfo{year}{1991}.
\bibitem[{{O'Sullivan} and {Downes}(2006)}]{osullivan06}
\bibinfo{author}{S.~{O'Sullivan}}, \bibinfo{author}{T.P. {Downes}},
  \bibinfo{journal}{\mnras} \bibinfo{volume}{366} (\bibinfo{year}{2006})
  \bibinfo{pages}{1329--1336}.
\bibitem[{{Pons} and {Geppert}(2007)}]{pons07}
\bibinfo{author}{J.A. {Pons}}, \bibinfo{author}{U.~{Geppert}},
  \bibinfo{journal}{\aap} \bibinfo{volume}{470} (\bibinfo{year}{2007})
  \bibinfo{pages}{303--315}.
\bibitem[{{Pons} and {Geppert}(2010)}]{pons10}
\bibinfo{author}{J.A. {Pons}}, \bibinfo{author}{U.~{Geppert}},
  \bibinfo{journal}{\aap} \bibinfo{volume}{513} (\bibinfo{year}{2010})
  \bibinfo{pages}{L12}.
\bibitem[{{Pons} et~al.(2009){Pons}, {Miralles} and {Geppert}}]{pons09}
\bibinfo{author}{J.A. {Pons}}, \bibinfo{author}{J.A. {Miralles}},
  \bibinfo{author}{U.~{Geppert}}, \bibinfo{journal}{\aap} \bibinfo{volume}{496}
  (\bibinfo{year}{2009}) \bibinfo{pages}{207--216}.
\bibitem[{{Shabaltas} and {Lai}(2012)}]{shabaltas12}
\bibinfo{author}{N.~{Shabaltas}}, \bibinfo{author}{D.~{Lai}},
  \bibinfo{journal}{\apj} \bibinfo{volume}{748} (\bibinfo{year}{2012})
  \bibinfo{pages}{148}.
\bibitem[{{Taflove} and {Brodwin}(1975)}]{taflove75}
\bibinfo{author}{A.~{Taflove}}, \bibinfo{author}{M.E. {Brodwin}},
  \bibinfo{journal}{IEEE Trans. Microwave Theory and Techniques}
  \bibinfo{volume}{23} (\bibinfo{year}{1975}) \bibinfo{pages}{623--630}.
\bibitem[{{Toro}(2009)}]{Toro}
\bibinfo{author}{E.~{Toro}}, \bibinfo{title}{Riemann Solvers and Numerical
  Methods for Fluid Dynamics}, \bibinfo{publisher}{Springer},
  \bibinfo{year}{2009}.
\bibitem[{{T{\'o}th} et~al.(2008){T{\'o}th}, {Ma} and {Gombosi}}]{toth08}
\bibinfo{author}{G.~{T{\'o}th}}, \bibinfo{author}{Y.~{Ma}},
  \bibinfo{author}{T.I. {Gombosi}}, \bibinfo{journal}{Journal of Computational
  Physics} \bibinfo{volume}{227} (\bibinfo{year}{2008})
  \bibinfo{pages}{6967--6984}.
\bibitem[{{Tran} and {Dritschel}(2010)}]{tran10}
\bibinfo{author}{C.V. {Tran}}, \bibinfo{author}{D.G. {Dritschel}},
  \bibinfo{journal}{Physics of Fluids} \bibinfo{volume}{22}
  (\bibinfo{year}{2010}) \bibinfo{pages}{037102}.
\bibitem[{{Vainshtein} et~al.(2000){Vainshtein}, {Chitre} and
  {Olinto}}]{vainshtein00}
\bibinfo{author}{S.I. {Vainshtein}}, \bibinfo{author}{S.M. {Chitre}},
  \bibinfo{author}{A.V. {Olinto}}, \bibinfo{journal}{Phys. Rev. E}
  \bibinfo{volume}{61} (\bibinfo{year}{2000}) \bibinfo{pages}{4422--4430}.
\bibitem[{{van Leer}(1977)}]{vanleer77}
\bibinfo{author}{B.~{van Leer}}, \bibinfo{journal}{Journal of Computational
  Physics} \bibinfo{volume}{23} (\bibinfo{year}{1977}) \bibinfo{pages}{276}.
\bibitem[{{Yee}(1966)}]{yee66}
\bibinfo{author}{K.S. {Yee}}, \bibinfo{journal}{IEEE Trans. on Antennas and
  Propagat.}  (\bibinfo{year}{1966}) \bibinfo{pages}{302--307}.

\end{thebibliography}

\end{document}